\documentclass[AMA,STIX1COL]{WileyNJD-v2}
\articletype{Research Article}
\usepackage[english]{babel}
\usepackage{arydshln}
\usepackage{amsmath}
\usepackage{amsthm}
\usepackage{amssymb, soul}
\usepackage{mathtools}
\usepackage{bm}
\usepackage{enumitem}
\usepackage{setspace}
\usepackage{threeparttable}
\usepackage{tabularx}
\usepackage{booktabs}
\usepackage{longtable}
\usepackage{array}
\usepackage{graphicx}

\setcitestyle{numbers,super,sort&compress}

\usepackage{lscape}

\usepackage{tikz}
\usetikzlibrary{arrows.meta, positioning, shadows.blur, calc}

\usepackage{algorithm}
\usepackage{algpseudocode}

\definecolor{ForestGreen}{RGB}{34,139,34}

\algnewcommand{\Input}{\item[\textbf{Input:}]}
\algnewcommand{\Output}{\item[\textbf{Output:}]}

\begin{document}

\title{
The FORSS Framework for Sample Size and Power Calculations With Win Statistics for Hierarchical Endpoints
}

\author[1,2]{Baoshan Zhang}

\author[1,2]{Huiman Barnhart*}

\author[1]{Yuan Wu}

\author[1,2]{Roland A. Matsouaka} 

\authormark{Baoshan Zhang {et al.}}

\address[1]{\orgdiv{Department of Biostatistics and Bioinformatics}, \orgname{Duke University}, \orgaddress{\city{Durham}, \state{North Carolina}, \country{USA}}}
\address[2]{\orgdiv{Duke Clinical Research Institute} \orgname{Duke University Medical Center}, \orgaddress{\city{Durham}, \state{North Carolina}, \country{USA}}}

\corres{*\email{huiman.barnhart@duke.edu}}

\presentaddress{2424 Erwin Rd, Durham, NC 27705}

\abstract{
Win statistics have gained increasing popularity as primary analysis methods for clinical trials with hierarchical endpoints (HEs) as primary endpoints. However, existing sample size and power calculation approaches in trial design still face several limitations and challenges: simulation-based approaches are computationally intensive, while existing formula-based methods often rely on simplifying assumptions such as independence among HEs, or require specification of overall win statistics and tie probability that are difficult to elicit a priori in practice. To address these challenges, we propose the FORSS framework, a FORmula-based Super-Sample approach that allows investigators to specify marginal treatment effects using familiar metrics (e.g., hazard ratios, mean differences, and risk differences) together with a flexible joint working distribution for the HEs. Rather than repeatedly simulating full trials at each candidate sample size, FORSS uses super-samples to estimate the population-level plug-in quantities required by analytical formulas for both power and sample size calculation. We evaluated the performance of the proposed FORSS through extensive simulation studies. The results show that the formula-based FORSS closely matches empirical power across a wide range of scenarios while maintaining Type~I error rates near the nominal 5\% level. An illustration based on the HEART-FID trial further shows that endpoint-dependence specifications can materially affect projected power and required sample size when planning trials with HEs.
}

\maketitle

\section{Introduction}\label{introduction}
Composite endpoints are widely used in cardiovascular trials to assess multiple clinically important outcomes simultaneously\cite{montori_validity_2005,walker2024composite}. However, conventional time-to-first-event methods consider only time-to-event endpoints, focus on the first occurring event (even if of lesser importance), and treat all events equally. They ignore any subsequent events as well as the differences in severity (or clinical importance) that may exist across the component endpoints. This can thus mask the effects of treatment on critical, often-later-occurring events such as mortality\cite{neaton_key_2005, ferreira2007problems}. Against this backdrop of time-to-first-event endpoints, the use of prioritized, hierarchical endpoints (HEs) has gained increased attention as an analytical approach to evaluate composite endpoints while preserving the clinical importance of each endpoint.

To address these issues, a hierarchical-endpoint approach ranks  component endpoints by prioritizing events and endpoints according to their clinical importance or severity\cite{finkelstein_combining_1999}. Within this framework, win measures (such as the win ratio, win odds, net benefit, or desirability of outcome ranking (DOOR)) serve as the primary effect size metrics to quantify treatment benefits\cite{fandino2025efficient, pocock_win_2012, mao_defining_2024, barnhart_sample_2025}. The win measures can accommodate diverse endpoint types, including time-to-event, binary, ordinal, and continuous outcomes, while respecting clinical priorities.\cite{redfors_win_2020, dong_win_2023} The use of the win ratio as a measure of treatment effect has gained increased popularity across several clinical areas, particularly in cardiology\cite{gregson2025hierarchical, pocock2024win}. It  has been used in trials such as ATTR-ACT\cite{maurer_tafamidis_2018}, COAPT\cite{stone_transcatheter_2018}, and HEART-FID\cite{mentz_randomized_2021}. In these studies, win measures provided a more comprehensive summary of treatment benefit, compared to the traditional hazard ratio of a time-to-event composite endpoint, by integrating more clinical information. This shift is further supported by regulatory agencies, such as the Food and Drug Administration (FDA), which now recognize the win ratio as a valuable analytical tool for evaluating HEs in drug and device approval pathways\cite{yu_sample_2022, fda_multiple_2022}. This growing reliance on win measures in late-phase trials requires practical statistical methods to guide trial design, particularly with regard to power and sample size calculations.

To determine sample size and power at the design stage, investigators have predominantly relied on simulation-based procedures\cite{redfors_win_2020, mentz_randomized_2021, james_dapagliflozin_2024, barnhart_sample_2025, kondo_hierarchical_2024}. In a typical workflow, a joint distribution for the HEs is specified separately for the treatment and control arms, based on chosen parameters; trial datasets are generated under an initial total sample size $N$; a selected win measure and corresponding test statistics are computed; and the entire process is replicated thousands of times via Monte Carlo to obtain an empirical power. Since power depends on the total sample size $N$, investigators must iteratively adjust the sample size and re-run the full Monte Carlo simulation at each step until the target statistical power is reached. This iterative, trial-and-error process can be computationally intensive; the burden rises sharply with the number of HEs and, for a fixed number of replications, grows approximately quadratically with each candidate sample size $N$\cite{yu_sample_2022, zhou_calculating_2022}. For example, Barnhart et al.\cite{barnhart_sample_2025} reported that approximately 14 days were required to evaluate empirical power with 5,000 simulation replicates, and this was for only a single fixed sample size. In practice, sample size calculations are typically performed across a spectrum of design scenarios and paramters, with varying endpoint configurations and treatment effects, making simulation-based procedures prohibitively time-consuming for exploring and comparing different design scenarios. As a result, fully simulation-based planning is often tedious (if not impractical) when multiple design scenarios must be explored.

An alternative to time-consuming simulation studies is a formula-based approach, which can deliver power and sample size calculations with far greater computational efficiency. Gasparyan et al. \cite{gasparyan_power_2021} developed a power and sample size formula for the win odds under a single ordinal outcome, assuming the absence of ties. Zhou et al.\cite{zhou_calculating_2022} proposed a formula-based derivation that relies on specific marginal distributional assumptions. However, their method is limited to HEs with exactly two endpoints and restricted to specific combinations of data types: (i) a time-to-event endpoint followed by a continuous or count endpoint, or (ii) a binary endpoint followed by a continuous or count endpoint. Their approach, formulated for the Finkelstein--Schoenfeld test\cite{finkelstein_combining_1999}, requires solving a high-order polynomial equation and does not readily generalize to HEs with more than two components or other data-type combinations. Mao et al.\cite{mao_sample_2022} provided a sample size formula involving the ``standard rank deviation,'' which requires numerical evaluation via Monte Carlo integration. Moreover, their approach applies separately to partially ordered outcomes and composite time-to-event outcomes, but not to their mixture, and no user-friendly software implementation is currently available. Yu and Ganju\cite{yu_sample_2022} derived a general closed-form sample size formula for the win ratio, assuming homogeneity of variances (estimated under the null hypothesis) and requiring two specific inputs: the overall win ratio and the overall probability of ties. However, these two quantities are rarely known and cannot be easily elicited from preliminary data, literature, or clinical investigators at the design stage, as HEs often combine disparate endpoints of different types, effect sizes, and follow-up horizons. In contrast, clinicians are much better at specifying marginal treatment effects, such as hazard ratios or mean differences, which are often reported in the literature or can be determined using preliminary data. This reality requires the derivation of the overall win measure from such marginal parameters for designing trials with HEs. 

Building on the work of Yu and Ganju\cite{yu_sample_2022}, Barnhart et al.\cite{barnhart_sample_2025} provided formula-based sample size and power calculations for several win measures, by linking overall win measures and the overall probability of ties to marginal win measures and marginal tie probabilities under an independence assumption. While their numerical evaluations suggested that correlation exerted minimal influence on power in the simulation scenarios they examined, Verbeeck et al.\cite{verbeeck_generalized_2019} demonstrated that correlation among HEs can fundamentally alter the effective contribution of lower-priority endpoints. These considerations motivate a flexible formula-based approach that permits both unequal variances and endpoint dependence to be incorporated explicitly at the design stage through a flexible joint working distribution. To this end, we propose the \emph{FORSS}  (FORmula-based Super Sample) framework for sample size and power calculation with win statistics for HEs.

The central idea of FORSS is to retain the computational efficiency of formula-based power and sample size calculations, while using super-sample-based estimation of the population-level plug-in quantities that related formulas require but that are generally unavailable in closed form for HEs. Rather than iteratively simulating numerous clinical trial data sets at each candidate sample size to evaluate empirical power, FORSS generates super samples with a large sample size from the specified joint distribution to estimate the quantities (independent of the super sample size) required by the analytical power and sample size formulas. This avoids the repetitive sample-size search that is typical of common simulation-based planning and thus substantially reduces computational burden. In this sense, FORSS uses simulation to estimate formula inputs, not to evaluate empirical power repeatedly across candidate sample sizes.

FORSS allows investigators to specify treatment effects through familiar marginal parameters, such as hazard ratios, mean differences, or risk differences, rather than requiring direct specifications of overall win measures and overall tie probabilities as design inputs. FORSS then connects these marginal specifications with the endpoint-dependence specifications to obtain the overall win measures required for design through a marginal-to-overall specification step. To incorporate the dependence among different component endpoints, FORSS uses a flexible joint working distribution. In this paper, we use copulas as one convenient implementation, but the framework is not restricted to any particular dependence structures. Therefore, the contribution of FORSS is not a new copula construction per se, nor merely a re-expression of overall win measures in terms of marginal quantities, but rather a general design framework that combines familiar marginal inputs, explicit endpoint-dependence specifications, and super-sample-based estimation of plug-in quantities for direct formula-based power and sample size calculations.

The remainder of this paper is organized as follows. Section~\ref{sec:formula-based-ssp} derives the sample size and power formulas based on win measures. Section~\ref{sec:FORSS-framework} describes the FORSS framework in detail. Section~\ref{sec:simulation} presents simulation studies evaluating the performance of the proposed methods. Section~\ref{sec:HEART-FID} illustrates the approach with an application using data from the HEART-FID trial as preliminary data for future trial design. Section~\ref{sec:discussion} concludes with a discussion.

\section{Formula-based Sample Size and Power Calculation}\label{sec:formula-based-ssp}

\subsection{Win Statistics}\label{basic-framework}

We adopt the U-statistics framework of Bebu and Lachin \cite{bebu_large_2016} and Dong et al. \cite{dong_win_2023} for HEs with mixed data types. We introduce the generic U-statistic framework underlying both observed-data inference and the design-stage development that follows. Suppose there are $m$ and $n$ subjects assigned to the treatment and control arms, respectively, with the allocation ratio $r = n/m$ and total sample size $N = m + n$. 
Let $\mathcal{D}$ be an HE  that consist of $Q$ endpoints observed over a follow-up period $S$, where the component endpoints in $\mathcal{D}$ are ranked by clinical importance (or severity) and may differ by their type (e.g., time-to-event, binary, continuous, count, or ordinal). 

For a subject $i$ in the treatment arm ($i = 1, \ldots, m$), we denote by $\mathcal{D}_{i}^{(T)} = \{ D_{1i}^{(T)}, D_{2i}^{(T)}, \ldots, D_{Qi}^{(T)}; S \}$, the observed HE where the components ordered by severity or  clinical importance. For a non-time-to-event component, $D_{qi}^{(T)}$ denotes the observed value of the 
$q$th component. When the $q$-th outcome is a time-to-event endpoint, let $T_{qi}^{(T)}$ denote the true event time. In the fixed follow-up setting considered in this paper, $S$ serves as a common administrative censoring time and the observed time-to-event component is $D_{qi}^{(T)} = \min\{T_{qi}^{(T)}, S\},$ and $ \delta_{qi}^{(T)} = I(T_{qi}^{(T)} \leq S),
$ where $\delta_{qi}^{(T)}$ is the event indicator and $I(\cdot)$ is the indicator function. 
Thus, time-to-event components are compared up to the common design horizon $S$.
Similarly, for a subject $j$ in the control arm ($j=1,\ldots,n$), we define 
$\mathcal{D}_{j}^{(C)} = \{D_{1j}^{(C)}, D_{2j}^{(C)}, \ldots, D_{Qj}^{(C)}; S\}$ 
with the corresponding control-arm quantities.

To estimate the win measures, pairwise comparisons are performed between each subject $i$ in the treatment arm and every subject $j$ in the control arm. Starting with the highest-priority outcome $D_{1}$, a pair of subjects is compared to determine which subject has the more favorable outcome (i.e., the winner). If there is no winner or the pair cannot be compared on $D_{1}$, the comparison proceeds to the next outcome $D_2$. This process continues sequentially throughout the hierarchy $\mathcal{D}$ until a winner is determined or we reach the last endpoint $D_{Q}$ without a winner. In the latter scenario, the pair is declared a tie. The indicators of winning, losing, and tie  for each outcome $q$ are defined as $W_q^{ij}, L_q^{ij},$ and $\Omega_q^{ij}$, respectively, according to the data types specified in Table~\ref{tab:comparison_rules}.

\begin{table}[ht]
\begin{threeparttable}
\caption{Pairwise comparison rules for winner, loser, and tie indicators by outcome data type.}
\label{tab:comparison_rules}
\begin{center}
\begin{tabular}{lllll}
\toprule
\textbf{Outcome Data Type} & \textbf{Winning $\left(W_q^{ij}\right)$} & \textbf{Losing $\left(L_q^{ij}\right)$} & \textbf{Tie $\left(\Omega_q^{ij}\right)$} \\
\midrule
Binary & $I\left(D_{qi}^{(T)} > D_{qj}^{(C)}\right)$ & $I\left(D_{qi}^{(T)} < D_{qj}^{(C)}\right)$ & $I\left(D_{qi}^{(T)} = D_{qj}^{(C)}\right)$ \\
\addlinespace
Continuous, Ordinal, or Count & $I\left( D_{qi}^{(T)} > D_{qj}^{(C)} + \delta_q \right)$ & $I\left(D_{qi}^{(T)} < D_{qj}^{(C)} - \delta_q\right)$ & $I\left(\big|D_{qi}^{(T)} - D_{qj}^{(C)}\big| \leq \delta_q\right)$ \\
\addlinespace
Time-to-event & $\delta_{qj}^{(C)} I\left(D_{qi}^{(T)} > D_{qj}^{(C)} +  \delta_q\right)$ & $\delta_{qi}^{(T)} I\left(D_{qi}^{(T)} < D_{qj}^{(C)}-   \delta_q\right)$ & $\left(1 - W_q^{ij}\right)\left(1 - L_q^{ij}\right)$ \\
\bottomrule
\end{tabular}
\begin{tablenotes}\footnotesize
    \item $D_{qi}^{(T)}$ and $D_{qj}^{(C)}$ represent the $q$-th outcome for treatment subject $i$ and control subject $j$, respectively. 
    $I(A)$ is the indicator function, i.e., $I(A)=1$ if condition $A$ is satisfied; otherwise $I(A)=0$. 
    The constant $\delta_q \geq 0$ is a pre-specified clinically meaningful threshold needed to declare a win when comparing on the $q$-th outcome.
\end{tablenotes}
\end{center}
\end{threeparttable}
\end{table}

Based on the pairwise comparison indicators provided in Table~\ref{tab:comparison_rules}, we define the win and loss kernel functions, which aggregate information across all $Q$ endpoints:
\begin{equation}
\varphi_{w}^{ij} = W_{1}^{ij} + \sum_{q = 2}^{Q} \left( \prod_{p = 1}^{q - 1} \Omega_{p}^{ij} \cdot W_{q}^{ij} \right), \quad \varphi_{l}^{ij} = L_{1}^{ij} + \sum_{q = 2}^{Q} \left( \prod_{p = 1}^{q - 1} \Omega_{p}^{ij} \cdot L_{q}^{ij} \right) \label{eq:kernel_functions}
\end{equation}
The win statistic $U_{w}$ and loss statistic $U_{l}$ are then defined as averages across all $m \times n$ pairwise comparisons:
\begin{equation}
U_{\nu} = \frac{1}{mn} \sum_{i = 1}^{m} \sum_{j = 1}^{n} \varphi_{\nu}^{ij}, \quad \nu \in \{w, l\} \label{eq:u_statistics}
\end{equation}
The tie statistic is defined as $U_{\Omega} = 1 - (U_{w} + U_{l})$. Accordingly, each pairwise comparison results in exactly one of three outcomes: a win, a loss, or a tie. We denote the corresponding population probabilities as $\tau_{w}$, $\tau_{l}$, and $\tau_{\Omega}$, where $\tau_{w} + \tau_{l} + \tau_{\Omega} = 1$. The exact variances of the win or loss statistics are
\begin{equation}
\sigma_{\nu\nu}^2\coloneqq\mathrm{Var}(U_{\nu}) = \frac{n - 1}{mn}\xi_{\nu\nu}^{10} + \frac{m - 1}{mn}\xi_{\nu\nu}^{01} + \frac{1}{mn}\xi_{\nu\nu}^{11}, \quad \nu \in \{w, l\} \label{eq:variance_u}
\end{equation}
where $\xi_{uv}^{10} = \text{Cov}(\varphi_{u}^{ij}, \varphi_{v}^{ij'})$, $\xi_{uv}^{01} = \text{Cov}(\varphi_{u}^{ij}, \varphi_{v}^{i'j})$, and $\xi_{uv}^{11} = \text{Cov}(\varphi_{u}^{ij}, \varphi_{v}^{ij})$, where $i \neq i'$ and $j \neq j'$. A detailed derivation of formula~\eqref{eq:variance_u} is provided in Appendix~\ref{sec:Appendix_Var}. The superscripts `10', `01', and `11' indicate whether, for the functions $\varphi_{u}^{ij}$ and  $\varphi_{v}^{i'j'}$ inside the covariance, the treatment subjects (first position) and control subjects (second position) are the same (1) or distinct (0). Thus, we have `10' if $i=i'$ but $j\neq j'$, `01' when $i\neq i'$ while $j= j'$, and  `11' when both $i=i'$ and  $j= j'$. 

Similarly, the exact covariance between $U_{w}$ and $U_{l}$ is given by
\begin{equation}\label{eq:cov_u}
\sigma_{wl}^{2} \coloneqq \text{Cov}(U_{w}, U_{l}) = \frac{n - 1}{mn}\xi_{wl}^{10} + \frac{m - 1}{mn}\xi_{wl}^{01} + \frac{1}{mn}\xi_{wl}^{11}.
\end{equation}
For large samples, $\displaystyle \frac{1}{mn}\approx 0$,  $\displaystyle \frac{n-1}{n}\approx 1,$ and  $\displaystyle \frac{m-1}{m}\approx 1$. Therefore, 
\begin{equation}\label{eq:cov_simplify}
\sigma_{uv}^{2} \approx \frac{1}{m}\xi_{uv}^{10} + \frac{1}{rm}\,\xi_{uv}^{01}, \quad u,v \in \{w,l\}.
\end{equation}

Under standard regularity conditions, as the total sample size $N \to \infty$ with fixed allocation ratio $r$, the win statistics $(U_w, U_l)$ converge jointly to a bivariate normal distribution \cite{bebu_large_2016, lehmann_elements_2004, lee_u-statistics_1990}. This asymptotic result is fundamental for the development of test statistics and inference procedures. This generic U-statistics framework and its asymptotic properties have also been extended to sequential monitoring designs for ordinal outcomes\cite{zhang2025sequential, wu2025group, zhang_sequential_2024}. 

\subsection{Sample Size Calculation and Power Formulas Based on Win Measures}\label{ssp-measures}

In this subsection, we leverage the asymptotic results derived in the previous subsection to present the power and sample size formulas for four win measures: the win ratio (WR), the net benefit (NB), the win odds (WO), and the desirability of outcome ranking (DOOR). 

To clarify the inputs required for the formulas, we distinguish between investigator-specified design inputs and population-level plug-in quantities, as summarized in Table \ref{tab:quantity_roles}. The intended treatment-arm sample size $m$ (with $n = rm$), the targeted power level $1-\beta$, the allocation ratio $r$, and the Type I error $\alpha$ are design inputs that can be chosen by the investigator. For power calculation, treatment-arm sample size $m$ with $n=rm$ is specified to obtain $1-\beta$. For sample size calculation, power $1-\beta$ is specified to obtain $m$ with $n=rm$. 

\begin{table}[htbp]
\centering
\caption{Roles of investigator-specified design inputs and population-level plug-in quantities in the FORSS framework.}
\label{tab:quantity_roles}
\begin{tabular}{lll}
\toprule
\textbf{Category} & \textbf{Parameters} & \textbf{Role in Trial Design} \\
\midrule
Design Inputs & $m, r, \alpha, \beta$ & Prespecified by investigator to define target design. \\
Population Quantities & $\tau_w, \tau_l, \tau_\Omega, \xi_{uv}^{10}, \xi_{uv}^{01}, \xi_{uv}^{11}$ & Estimated via FORSS; independent of design inputs. \\
\bottomrule
\end{tabular}
\end{table}
By contrast, the win, loss and tie probabilities ($\tau_w$, $\tau_l$, $\tau_\Omega$), and the covariance components ($\xi_{uv}^{10}, \xi_{uv}^{01}, \xi_{uv}^{11}$) are population-level plug-in quantities that must be estimated. Importantly, these population quantities do not depend on the investigator-specified design inputs. This separation is central to our framework because it allows the population-level plug-in quantities to be estimated within FORSS (Section~\ref{sec:FORSS-framework}) separately from the final power and sample size calculations.

All population quantities depend on the underlying hypothesis $H \in \{H_0, H_A\}$. We write $\tau_w(H)$, $\tau_l(H)$, $\tau_\Omega(H)$, and $\xi_{uv}^{rs}(H)$ to make this dependence explicit. Under $H_0$, $\tau_w(H_0) = \tau_l(H_0)$, so that $\mathrm{WR} = 1$, $\text{NB} = 0$, $\text{WO} = 1$, and $\text{DOOR} = 0.5$. For brevity, we suppress the hypothesis argument when the context is unambiguous.

\subsubsection{Large-Sample Variance and Test Statistics}
Building on the population quantities $\tau_w$, $\tau_l$, $\tau_\Omega$, $\xi_{uv}^{10}$, $\xi_{uv}^{01}$, and $\xi_{uv}^{11}$, we use the log(WR) to illustrate the derivation of the large-sample power and sample size formulas. Applying the delta method to $\log(\widehat{\mathrm{WR}})=\log(U_w/U_l)$ gives
\[
\mathrm{Var}_H\{\log(\widehat{\mathrm{WR}})\}\approx \frac{1}{m}\mathcal{A}_{\mathrm{WR}}(H),
\]
where
\[
\mathcal{A}_{\mathrm{WR}}(H)
=
\frac{\xi_{ww}^{10}(H)+\frac{1}{r}\xi_{ww}^{01}(H)}{\tau_w(H)^2}
+
\frac{\xi_{ll}^{10}(H)+\frac{1}{r}\xi_{ll}^{01}(H)}{\tau_l(H)^2}
-
2\frac{\xi_{wl}^{10}(H)+\frac{1}{r}\xi_{wl}^{01}(H)}{\tau_w(H)\tau_l(H)}.
\]
Analogous large-sample variance constants for the other win measures are summarized in Table~\ref{tab:power_functions}.

To test the null hypothesis $H_0:\mathrm{WR}=1$ of no treatment effect against a two-sided alternative at level $\alpha$, we use the test statistic
\[
T_{\mathrm{WR}}=
\frac{\log(\widehat{\mathrm{WR}})}{\sqrt{\mathcal{A}_{\mathrm{WR}}(H_0)/m}},
\]
which is asymptotically standard normal under under the null hypothesis $H_0$. We reject $H_0$ if $|T_{\mathrm{WR}}|>Z_{1-\alpha/2}$. The remaining win measures in Table~\ref{tab:power_functions} are handled analogously, with each test statistic is formed by dividing the corresponding effect estimator by its standard error under the null hypothesis $H_0$. It's important to note that the large-sample variance quantity $\mathcal{A}(H)$ depends on the randomization ratio $r$, components in the effect size,  and covariances $\xi_{uv}^{kl}$ with  $u,v \in \{w,l\}$ and $k,l \in \{0,1\}$, but not dependent  on the sample size $m$ with $n=rm$. Therefore, these quantities can be estimated more precisely with super samples with a super sample size that is much greater than $N=m+n$.

\begin{table}[htbp]
\centering
\setlength{\tabcolsep}{10pt}
\begin{threeparttable}
\caption{Measure-specific quantities for the unified power~\eqref{eq:power_general} and sample size~\eqref{eq:sample_size_general} formulas.}
\label{tab:power_functions}
\renewcommand{\arraystretch}{1.8}
\begin{tabular}{p{2.5cm} p{3.3cm} p{8.5cm}}
\toprule
\textbf{Win Measure} & \textbf{Effect Size $\Delta$} & \textbf{Large-Sample Variance Quantity $\mathcal{A}(H)$} \\
\midrule

$\log(\mathrm{WR})$
&
$\left|\log\!\left(\dfrac{\tau_w}{\tau_l}\right)\right|$
&
$\mathcal A_{\mathrm{WR}}(H)=\displaystyle
\frac{\xi_{ww}^{10}+\frac{1}{r}\xi_{ww}^{01}}{\tau_w^2}
+
\frac{\xi_{ll}^{10}+\frac{1}{r}\xi_{ll}^{01}}{\tau_l^2}
-
2\frac{\xi_{wl}^{10}+\frac{1}{r}\xi_{wl}^{01}}{\tau_w\tau_l}
$
\\[4pt]

$\mathrm{NB}$
&
$\left|\tau_w-\tau_l\right|$
&
$\mathcal A_{\mathrm{NB}}(H)=
\displaystyle \xi_{ww}^{10}+\frac{1}{r}\xi_{ww}^{01}
+
\xi_{ll}^{10}+\frac{1}{r}\xi_{ll}^{01}
-
2\left(\xi_{wl}^{10}+\frac{1}{r}\xi_{wl}^{01}\right)
$
\\[4pt]

$\log(\mathrm{WO})$
&
$\left|\log\!\left(\dfrac{\tau_w+\frac{1}{2}\tau_\Omega}{\tau_l+\frac{1}{2}\tau_\Omega}\right)\right|$
&
$\mathcal A_{\mathrm{WO}}(H)= \displaystyle\frac{4\,\mathcal A_{\mathrm{NB}}(H)}
{\left[1-\{\tau_w-\tau_l\}^2\right]^2}
$
\\[4pt]

$\mathrm{DOOR}$
&
$\displaystyle \frac{1}{2}\left|\tau_w-\tau_l\right|$
&
$\mathcal A_{\mathrm{DOOR}}(H)=
\displaystyle \frac{1}{4}\mathcal A_{\mathrm{NB}}(H)
$
\\[4pt]

\bottomrule
\end{tabular}

\begin{tablenotes}\footnotesize
\item[(1)] For readability, the hypothesis argument $H$ is suppressed in the table. The effect size $\Delta$ is evaluated under $H_A$, whereas the large-sample variance constants are evaluated under either $H_0$ or $H_A$ as required in formulas~\eqref{eq:power_general} and~\eqref{eq:sample_size_general}.
\item[(2)] At the trial planning stage, these population quantities are replaced by plug-in estimates from the FORSS framework (Section~\ref{sec:FORSS-framework}).
\item[(3)] NB and DOOR yield identical sample sizes because $\widehat{\mathrm{DOOR}}=0.5+0.5\,\widehat{\mathrm{NB}}$, implying $\mathcal A_{\mathrm{DOOR}}(H)=\frac{1}{4}\mathcal A_{\mathrm{NB}}(H)$ and $\Delta_{\mathrm{DOOR}}=\frac{1}{2}\Delta_{\mathrm{NB}}$, so that $\mathcal A(H)/\Delta^2$ is unchanged.
\end{tablenotes}
\end{threeparttable}
\end{table}

\subsubsection{Power and Sample Size Formulas} \label{subsubsec:ss-formula}

The power for a two-sided test shares a unified form across all four measures. Let $\Delta$ denote a measure-specific effect size under $H_A$ and let $\mathcal A(H)$ denote its corresponding large-sample variance quantity under hypothesis $H$, as defined in Table~\ref{tab:power_functions}. Then,
\begin{equation}\label{eq:power_general}
1 - \beta \approx \Phi \left( \frac{-z_{1-\alpha/2}\sqrt{\mathcal{A}({H_0})} + \sqrt{m}\,\Delta}{\sqrt{\mathcal{A}({H_A})}} \right),
\end{equation}
where $\Phi(\cdot)$ is the cumulative distribution function of the standard normal distribution. Detailed steps leading to the formula \eqref{eq:power_general}, for each win measure, are provided in Appendix~\ref{Sec:Appendix_PowerCalc}.

By inverting equation~\eqref{eq:power_general}, we obtain a concise closed-form formula for the required treatment-arm sample size $m$, i.e.,
\begin{equation}\label{eq:sample_size_general}
m = \frac{\left(z_{1-\alpha/2}\,\sqrt{\mathcal{A}({H_0})} 
+ z_{1-\beta}\,\sqrt{\mathcal{A}({H_A})}\right)^2}{\Delta^2},
\end{equation}
which corresponds to the total sample size of $N = (1+r) m$. The closed-form formula~(\ref{eq:sample_size_general}) relies on the large-sample simplifications in equation (\ref{eq:cov_simplify}). Alternatively, the power function can be evaluated or inverted numerically using the exact finite-sample variances (retaining the $\xi_{uv}^{11}(H)$ terms 
from formula~ \eqref{eq:variance_u}) without the large-sample simplification. This approach is implemented in our software. 

\section{The FORSS Framework}\label{sec:FORSS-framework}

Like our proposed formulas, most established formula-based methods for power calculation require direct, a priori specification of an overall treatment effect expressed in terms of win measures, such as a target WR $=\tau_{w}/\tau_{l}$ or NB $=\tau_{w} - \tau_{l}$, and the overall probability of ties $\tau_{\Omega}$ that are not readily available in practice~\cite{yu_sample_2022, gasparyan_power_2021, mao_sample_2022}. Similarly, our proposed formulas, as shown in equation~\eqref{eq:cov_simplify}--\eqref{eq:sample_size_general} and Table~\ref{tab:power_functions}, also require additional specifications in $\Delta$ and large-sample variance constant $\mathcal{A}(H)$ under both $H_{0}$ and $H_{A}$. These quantities are usually not readily available from literature, preliminary data, or the clinical investigators to be used by the applied statisticians at the study planning stage.

This challenge arises from a fundamental disconnect between the required parameters and available clinical evidence. Clinicians and existing literature seldom quantify efficacy using win measure metrics for the HE needed in the planned study. Instead, treatment effects are often conceptualized and reported on at the marginal level measures of individual component endpoints of an HE such as the hazard ratios for time-to-event outcomes, the mean differences for continuous measures, and the risk differences for binary events. Consequently, asking investigators to provide a reliable overall win measure or the probability of ties is neither intuitive nor practical. Furthermore, reliance on a single win measure metric obscures how each component of the HE contributes to the overall treatment effect, whereas transparency regarding the contribution of each component is increasingly emphasized by regulatory agencies~\cite{fda_patient_2023}.

Barnhart et al.~\cite{barnhart_sample_2025} addressed part of this gap by proposing a method to derive overall win measures from marginal treatment effects. Their approach allows investigators to specify marginal outcome distributions under both $H_0$ and $H_A$, thereby connecting the composite effect to clinically interpretable parameters. To simplify the derivation, they adopted an independence assumption across endpoints. While this assumption enhances tractability, it may not reflect the fact that correlations between the different component endpoints of an HE are commonly observed in clinical practice. For example, endpoints such as mortality, hospitalization, and functional status are often related and thus their correlations should not be ignored. While Barnhart et al.~\cite{barnhart_sample_2025} showed that the correlation seems to have minimal impact on powers or sample sizes based on the examples that they investigated, we believe that it is important to have a more in depth examination of such impact. Therefore, we propose the \emph{FORmula-based Super Sample} (FORSS) framework, which builds on specifications of marginal effect sizes while relaxing the independence assumption. 

The FORSS framework offers three key advantages. First, it preserves the clinical interpretability of the marginal approach. The investigators can specify treatment effects using familiar parameters such as hazard ratios, mean differences, and risk differences. In this way, the efficacy of each component in the HE is directly reflected in the design assumptions. Second, FORSS allows the inherent dependence among the component endpoints to be incorporated through a flexible working joint distribution. In this paper, we use a copula-based construction as a convenient implementation, which allows trialists to incorporate realistic dependency patterns derived from pilot data or clinical judgment. Third, FORSS uses a super sample to estimate the population-level plug-in quantities, $\Delta$ and $\mathcal{A}(H)$, required by the power and sample size formulas. This computational role is fundamentally different from conventional simulation-based planning, which repeatedly simulates full trials at candidate sample sizes to estimate empirical power. This retains the flexibility of model-based data generation while reducing computational burden. 

\subsection{FORSS Framework Specification}\label{subsec:FORSS-specification}

\begin{figure}
\centering
\begin{tikzpicture}[
    node distance = 0.85cm and 1.6cm,
    >=Latex,
    stage/.style={
        rectangle,
        rounded corners=4pt,
        draw=blue!50!black,
        thick,
        align=left,
        text width=8.6cm,
        minimum height=1.45cm,
        inner sep=8pt,
        fill=blue!4
    },
    output/.style={
        rectangle,
        rounded corners=4pt,
        draw=blue!60!black,
        very thick,
        align=left,
        text width=8.6cm,
        minimum height=1.25cm,
        inner sep=8pt,
        fill=blue!10
    },
    sidebox/.style={
        rectangle,
        rounded corners=4pt,
        draw=black!65,
        thick,
        dashed,
        align=left,
        text width=5.7cm,
        inner sep=8pt,
        fill=gray!6
    },
    arrow/.style={
        -{Latex[length=3mm,width=2mm]},
        thick,
        draw=blue!60!black
    },
    sidearrow/.style={
        -{Latex[length=2.7mm,width=1.8mm]},
        thick,
        dashed,
        draw=black!65
    }
]

\node[stage] (stage1) {
\textbf{Stage 1: Marginal specification}\\
For each endpoint $q\in\{1,\ldots,Q\}$, specify endpoint type, $F_q^{(C)}$, and $\Delta_q$; these determine $F_q^{(T)}$.
};

\node[stage, below=of stage1] (stage2) {
\textbf{Stage 2: Specification of Joint Distributions of the HE}\\
Specify the joint dependence structure. For illustration, copula $C_{\bm{\mathcal{R}}}$ is used as the joint distribution incorporating specifications of the marginal distributions and the dependence structure.
};

\node[stage, below=of stage2] (stage3) {
\textbf{Stage 3: Super-sample estimation}\\
Generate super samples of size $N_{\textup{sp}}$ and estimate $\bar{\tau}_w$, $\bar{\tau}_l$, and $\bar{\xi}_{uv}^{rs}$ via Algorithm~\ref{alg:forss}.
};

\node[output, below=of stage3] (output) {
\textbf{Formula-based design output}\\
Plug the estimated quantities into analytical power and sample size formulas via Table \ref{tab:power_functions}.
};

\node[sidebox, right=2.0cm of stage2.east, anchor=west] (dep) {
\textbf{Dependence specification in practice}\\[3pt]
\textbf{No pilot data:} specify a plausible range of $\bm{\mathcal{R}}$ and perform sensitivity analysis.\\[6pt]
\textbf{Pilot data available:} calibrate $\bm{\mathcal{R}}$ to reproduce observed $\widehat{\bm{\mathcal{K}}}$.\\[6pt]
\textbf{High uncertainty:} allow blinded interim sample size re-estimation.
};

\draw[arrow] (stage1) -- (stage2);
\draw[arrow] (stage2) -- (stage3);
\draw[arrow] (stage3) -- (output);

\draw[sidearrow] (dep.west) -- ++(-0.5,0) |- (stage2.east);

\end{tikzpicture}
\caption{Overview of the FORSS framework and practical strategies for dependence specification.}
\label{fig:forss_overview}
\end{figure}
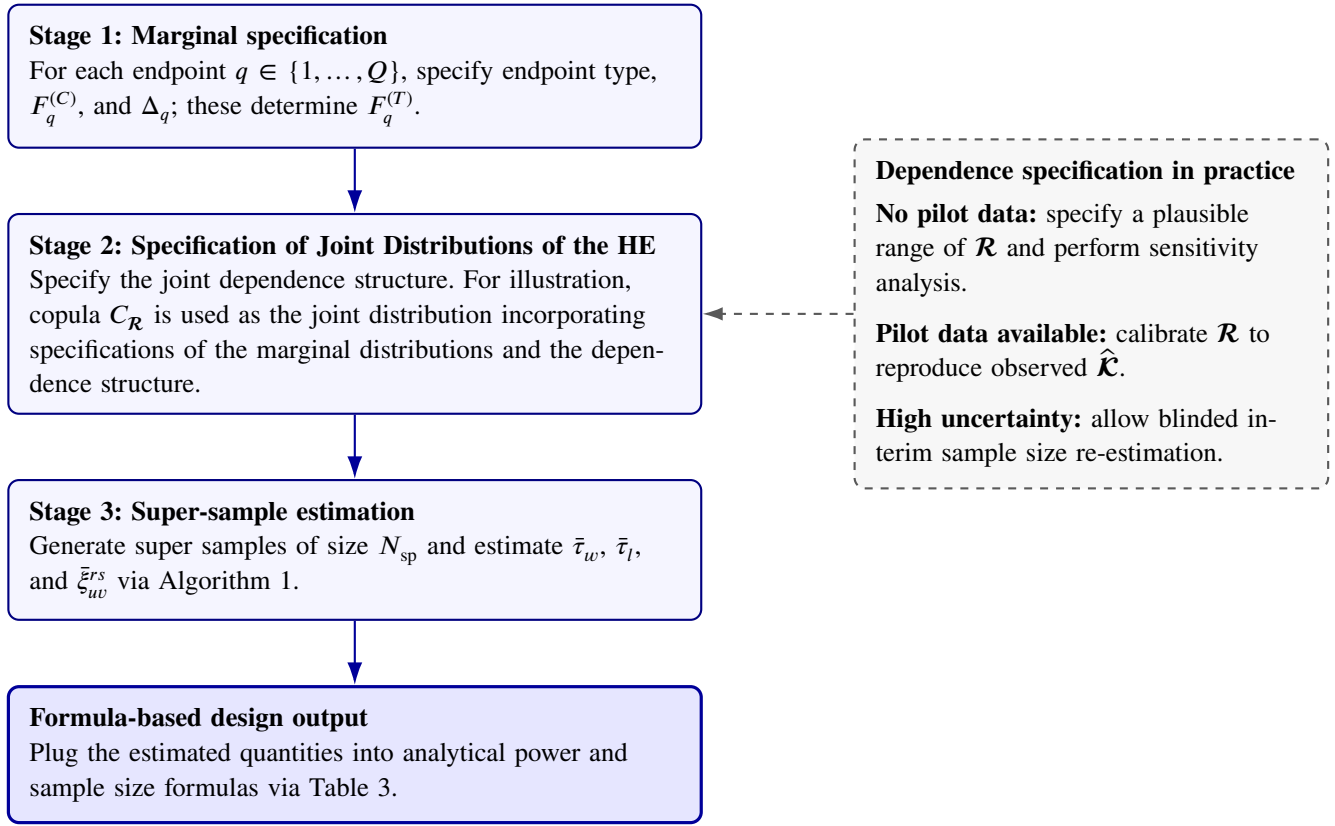

The FORSS framework consists of three stages: (i) marginal specification, (ii) joint distribution specification, and (iii) super-sample-based estimation of plug-in quantities. A schematic overview of the FORSS framework is shown in Figure~\ref{fig:forss_overview}.

\paragraph{Stage 1: Marginal Specification.} 
Consider $\mathcal{D} = (D_1, D_2, \ldots, D_Q)$ an HE consisting of $Q$ outcomes, ordered by clinical priority. For each outcome $q \in \{1, \ldots, Q\}$, the investigator specifies the data type, the marginal distribution under the control arm $F_q^{(C)}$, and the marginal treatment effect $\Delta_q$. From $F_q^{(C)}$ and $\Delta_q$, the treatment arm marginal distribution $F_q^{(T)}$ is then specified.

\paragraph{Stage 2: Specification of Joint Distributions of HE.}
To complete the design specification, FORSS requires a working joint distribution of the $Q$ outcomes to generate super-sample replicates. Let $\mathcal{M}$ denote a user-specified joint working specification which, together with the marginal distributions from Stage~1, induces the treatment- and control-arm joint distributions $\mathcal{F}^{(T)}$ and $\mathcal{F}^{(C)}$. In principle, the joint working distributions in FORSS may be specified by any meaningful multivariate distributions for the HE under consideration. In practice, a copula-based representation is convenient and is used here as the primary illustration. Compared with direct parametric joint distributions tied to specific outcome families, copula-based constructions provide a more natural default in FORSS because they preserve the investigator-specified marginal distributions by construction, accommodate endpoints of different data types, and allow the dependence structure to be introduced separately from the marginal models. This is particularly appealing here, because trial design in FORSS is explicitly anchored on marginal endpoint distributions and marginal treatment effects. 

By Sklar's theorem, any multivariate distribution can be represented in terms of its marginal distributions and a copula capturing the dependence structure~\cite{nelsen_introduction_2006}. Under this representation,
\begin{equation}\label{eq:joint_treatment}
\mathcal{F}^{(T)}(\bm d)
=
C_{\bm{\mathcal R}}\!\left(
F_1^{(T)}(d_1),\ldots,F_Q^{(T)}(d_Q)
\right),
\qquad \bm d=(d_1,\ldots,d_Q),
\end{equation}
and analogously for $\mathcal{F}^{(C)}$ using the control-arm marginals $\{F_q^{(C)}\}_{q=1}^Q$. When a copula working model $C_{\bm{\mathcal R}}$ is used, we assume a common copula family and dependence parameter matrix $\bm{\mathcal R}$ across both arms, reflecting the working assumption that treatment acts through the marginal distributions.

\paragraph{Stage 3: Super-Sample-Based Estimation of Plug-in Quantities.}
Given the joint distributions $\mathcal{F}^{(C)}$ and  $\mathcal{F}^{(T)}$ from Stage~2, FORSS estimates the win and loss probabilities $\tau_w$ and $\tau_l$, together with covariance components $\xi_{uv}^{rs}$, under both $H_0$ and $H_A$, as required by Table~\ref{tab:power_functions}. Closed-form expressions for these quantities are generally unavailable for HE because of mixed endpoint data types, the flexible marginal specification, and nontrivial dependence structures. Rather than approximating power through repeated simulations of full trial data at candidate sample sizes, FORSS generates $b$ super-sample replicates (to control the precision of the estimated quantities in Table \ref{tab:power_functions}), each of a large size $N_{\textup{sp}}$ per arm from the $\mathcal{F}^{(C)}$ and  $\mathcal{F}^{(T)}$. The super-sample replicates are then used to estimate the population-level plug-in quantities directly and then 
substitutes the resulting estimates into the analytical formulas~(\ref{eq:power_general})--(\ref{eq:sample_size_general}). Because a single-super sample may yield estimates with non-negligible Monte Carlo variability, we use the adaptive procedure shown in Algorithm~\ref{alg:forss}, which iteratively updates the plug-in estimates until the standard errors of the running averages for the $\tau$-quantities and the $\xi$-quantities fall below pre-specified tolerance levels $\varepsilon_{\tau}$ and $\varepsilon_{\xi}$, respectively. To achieve these pre-specified prevision, we may need $b$ copies of super samples (see Algorithm 1).

\paragraph{Estimating the population-level plug-in quantities.}
The power and sample size calculation formulas require $\tau_w(H)$, $\tau_l(H)$, $\tau_\Omega(H)$, and $\xi_{uv}^{kl}(H)$. Because these are population-level functionals, they do not depend on the planned sample sizes $(m,n)$ or allocation ratio $r$. We therefore estimate them in FORSS using balanced super samples with $N_{\textup{sp}}$, where $N_{\textup{sp}}$ controls only Monte Carlo precision. For one super-sample replicate, we propose the following consistent empirical estimators: 
\begin{align}
\widehat{\tau}_u
&=
\frac{1}{N_{\textup{sp}}^2}
\sum_{i=1}^{N_{\textup{sp}}}\sum_{j=1}^{N_{\textup{sp}}}\varphi_u^{ij}
\end{align}
and 
\begin{align}
\widehat{\xi}_{uv}^{10}
&=
\frac{1}{N_{\textup{sp}}^2(N_{\textup{sp}}-1)}
\sum_{i=1}^{N_{\textup{sp}}}
\sum_{\substack{j_1,j_2=1\\ j_1\neq j_2}}^{N_{\textup{sp}}}
\varphi_{uv}^{i;j_1,j_2}
-\widehat{\tau}_u\widehat{\tau}_v, \quad 
\widehat{\xi}_{uv}^{01}=
\frac{1}{N_{\textup{sp}}^2(N_{\textup{sp}}-1)}
\sum_{j=1}^{N_{\textup{sp}}}
\sum_{\substack{i_1,i_2=1\\ i_1\neq i_2}}^{N_{\textup{sp}}}
\varphi_{uv}^{i_1,i_2;j}
-\widehat{\tau}_u\widehat{\tau}_v,\quad
\widehat{\xi}_{uv}^{11}= 
\frac{1}{N_{\textup{sp}}^2}
\sum_{i=1}^{N_{\textup{sp}}}\sum_{j=1}^{N_{\textup{sp}}}
\varphi_{uv}^{i;j}
-\widehat{\tau}_u\widehat{\tau}_v.\nonumber
\end{align}
where
\[
\varphi_{uv}^{i_1,i_2;j}
=\tfrac12\!\left(\varphi_u^{i_1j}\varphi_v^{i_2j}+\varphi_u^{i_2j}\varphi_v^{i_1j}\right),\quad
\varphi_{uv}^{i;j_1,j_2}
=\tfrac12\!\left(\varphi_u^{ij_1}\varphi_v^{ij_2}+\varphi_u^{ij_2}\varphi_v^{ij_1}\right),\quad
\varphi_{uv}^{i;j}=\varphi_u^{ij}\varphi_v^{ij}.
\]
These replicate-specific estimates are computed under both $H_0$ and $H_A$ and then averaged adaptively in Algorithm~\ref{alg:forss}. Substituting these consistent estimators into quantities in Table \ref{tab:power_functions} yields consistent estimators for the population-level plug-in quantities $\Delta$ and $\mathcal{A}(H)$.

\begin{algorithm*}[ht!]
\caption{Adaptive super-sample estimation of plug-in quantities in FORSS}
\label{alg:forss}
\begin{spacing}{1.2}
\begin{algorithmic}

\Input Control-arm marginals $\{F_q^{(C)}\}_{q=1}^Q$; marginal treatment effects $\{\Delta_q\}_{q=1}^Q$; joint working model $\mathcal{M}$ for the $Q$ endpoints.
\Statex \hspace{\algorithmicindent} Super-sample size $N_{\text{sp}}$ per arm ; iteration limits $b_{\min}$ and $b_{\max}$; convergence thresholds $\varepsilon_{\tau}$ and $\varepsilon_{\xi}$.
\Output Stable estimates $\bar{\tau}_w(H)$, $\bar{\tau}_l(H)$, and $\bar{\xi}_{uv}^{rs}(H)$ for $u, v \in \{w, l\}$, $(r,s) \in \{(1,0), (0,1), (1,1)\}$, under both $H_0$ and $H_A$.
\Statex \rule{\linewidth}{0.4pt}
\State Derive treatment-arm marginals $F_q^{(T)}$ from $F_q^{(C)}$ and $\Delta_q$ for each $q$.
\State Construct the joint distributions $\mathcal{F}^{(T)}$ and $\mathcal{F}^{(C)}$ by combining the marginals with the user-specified joint working model $\mathcal{M}$.
\State Initialize $b \gets 0$, $\widehat{\mathrm{SE}}^{(\tau)}_{\max} \gets \infty$, and $\widehat{\mathrm{SE}}^{(\xi)}_{\max} \gets \infty$.

\While{$b < b_{\max}$ \textbf{and} $\left(\widehat{\mathrm{SE}}^{(\tau)}_{\max} > \varepsilon_{\tau} \textbf{ or } \widehat{\mathrm{SE}}^{(\xi)}_{\max} > \varepsilon_{\xi}\right)$}
    \State $b \gets b + 1$
    \State Generate three independent super samples of size $N_{\text{sp}}$:
    \State \hspace{1em} Control sample: $\{\mathcal{D}_j^{(C)}\}_{j=1}^{N_{\text{sp}}} \sim \mathcal{F}^{(C)}$
    \State \hspace{1em} Treatment sample under $H_A$: $\{\mathcal{D}_i^{(T)}\}_{i=1}^{N_{\text{sp}}} \sim \mathcal{F}^{(T)}$
    \State \hspace{1em} Treatment sample under $H_0$: $\{\mathcal{D}_k^{(C*)}\}_{k=1}^{N_{\text{sp}}} \sim \mathcal{F}^{(C)}$
    \State Under $H_A$, compare $\{\mathcal{D}_i^{(T)}\}$ versus $\{\mathcal{D}_j^{(C)}\}$ to compute $\hat{\tau}_w^{(b)}(H_A)$, $\hat{\tau}_l^{(b)}(H_A)$, and $\hat{\xi}_{uv}^{rs(b)}(H_A)$.
    \State Under $H_0$, compare $\{\mathcal{D}_k^{(C*)}\}$ versus $\{\mathcal{D}_j^{(C)}\}$ to compute $\hat{\tau}_w^{(b)}(H_0)$, $\hat{\tau}_l^{(b)}(H_0)$, and $\hat{\xi}_{uv}^{rs(b)}(H_0)$.
    \If{$b \ge b_{\min}$}
    \State Compute $\widehat{\mathrm{SE}}^{(\tau)}_{\max}$ and $\widehat{\mathrm{SE}}^{(\xi)}_{\max}$ as the maximum estimated standard errors of the running averages of all $\tau$-quantities and all $\xi$-quantities, respectively, across $H_0$ and $H_A$.
    \EndIf
\EndWhile
\State \textbf{Return:} Averaged plug-in estimates, for example, $\bar{\tau}_w(H)=\frac{1}{b}\sum_{b'=1}^{b}\hat{\tau}_w^{(b')}(H)$, and similarly for the remaining quantities.

\end{algorithmic}
\end{spacing}
\end{algorithm*}

\subsubsection{Remarks of the FORSS}\label{subsec:FORSS-remarks}

\paragraph{Alternative joint working distributions.}
Copulas are not required to use the FORSS framework. Any joint working model compatible with the Stage~1 marginal specification can be used. For example, one may directly posit a bivariate exponential distribution for time-to-event outcomes~\cite{luo2017weighted}, a multivariate normal distribution for continuous outcomes, or a multivariate Bernoulli distribution for binary outcomes. This flexibility in the choice of joint working models affords the FORSS framework its versatility.

\paragraph{Role of the super-sample size $N_{\textup{sp}}$}
The super-sample size $N_{\textup{sp}}$ should not be confused with the planned trial sample sizes $m$ and $n$. The quantities $\tau_w$, $\tau_l$, and $\xi_{uv}^{kl}$ in Table~\ref{tab:power_functions} are population-level and therefore do not depend on the planned sample size $m$ and $n$. Rather than generating data with planned sample size $m$ and $n$, one can get better estimation by using $N_{\textup{sp}}$ for large value of $N_{\textup{sp}}$, say 2,000, for the sample size in each arm.  $N_{\textup{sp}}$ serves only to control the Monte Carlo variability of their estimation within each adaptive iteration. In this sense, $N_{\textup{sp}}$ is a tuning parameter for the super-sample estimation step rather than a design parameter for the planned trial itself. 

Larger values of $N_{\textup{sp}}$ typically reduce per-iteration Monte Carlo variability but increase the computational and memory cost of each replicate, whereas smaller values reduce per-replicate cost but may require more adaptive replicates, that is, a larger $b$, to satisfy the stopping rule. In our numerical studies, $N_{\textup{sp}}=2{,}000$ was generally sufficient, with $b$ typically ranging from 600 to 3{,}000. A larger value, such as $N_{\textup{sp}}=8{,}000$, can reduce the number of adaptive replicates needed for convergence, but this reduction does not necessarily translate into shorter total runtime because each replicate is more computationally expensive. Thus, $N_{\textup{sp}}$ should be viewed as a tuning parameter that balances per-replicate precision against computational cost.

Although the adaptive FORSS procedure also generates repeated super samples, it differs from a conventional simulation-based power calculation. FORSS uses the super samples only to estimate the population-level plug-in quantities in Table~\ref{tab:power_functions}; it does not repeatedly simulate complete trials at candidate sample sizes and estimate empirical rejection probabilities. This distinction is the main source of computational savings relative to fully simulation-based sample size search.

\subsection{Practical Strategies for Endpoint-Dependence Specification in FORSS}\label{subsec:dependence-forss}

When the joint distribution $\mathcal{M}$ in FORSS is implemented through a copula, dependence is parameterized by the correlation matrix of an underlying latent vector, denoted by $\bm{\mathcal{R}}=(\rho_{qq'})_{1 \le q,q' \le Q}$, where $\rho_{qq}=1$ and $\rho_{qq'}=\rho_{q'q}$. However, in practice, trialists are more likely to have access to observed-scale summaries of association rather than latent correlations $\bm{\mathcal{R}}$. These are referred to as observed concordance summaries and denoted $\bm{\mathcal{K}} := \bigl(\kappa_{qq'}\bigr)_{1 \le q, q' \le Q}$, where concordance refers to a rank-based measure of monotonic association on the observed data scale. 

Under a Gaussian copula with continuous margins, a closed-form relationship exists between the latent correlation $\bm{\mathcal{R}}$ and certain concordance measures on the observed scale $\bm{\mathcal{K}}$~\cite{nelsen_introduction_2006}. For example, when Kendall's $\tau$ is used, then
\[
\rho_{qq'} = \sin\!\left(\frac{\pi}{2}\kappa_{qq'}\right).
\]
However, this relationship becomes substantially more complicated  once the HEs involve discrete outcomes, censoring, bounded endpoints, or mixed data types~\cite{genest2007primer, de2011copula}. In such settings, the relationship between the latent copula parameter and the observed concordance depends jointly on the marginal distributions $\{F_q^{(C)}\}_{q=1}^Q$ and on any censoring present in the component endpoints. Therefore, a universal closd-form transformation from observed $\bm{\mathcal{K}}$ to latent $\bm{\mathcal{R}}$ is generally generally unavailable, and the $\bm{\mathcal{R}}$ may instead be calibrated numerically to reproduce target observed concordance summaries $\bm{\mathcal{K}}$.

\begin{itemize}
\item {\bf Strategy 1: Direct specification with sensitivity analysis.}\label{Str:1}
When no pilot data are available for the planned population, which is often the more common case in practice, we recommend specifying a clinically plausible range of latent dependence values for $\bm{\mathcal{R}}$ and performing the FORSS calculation over that range. The resulting sample size can then be chosen on the conservative side, in the same spirit as standard sensitivity analyses over nuisance parameters in trial design. In this setting, the goal is not to identify a single ``correct'' dependence value, but rather to ensure sufficient power for a chosen sample size based on reasonable range of plausible dependence structures. If this is possible, an initial sample size should be chosen conservatively with a guess of the dependence structure and then update the sample size during interim blinded sample size re-estimation (see below for blinded sample size re-estimation).

\item {\bf Strategy 2: Pilot-data-based calibration.}\label{Str:2}
When pilot data or historical data for all components of the HEs are available, the observed concordance summaries may be estimated first, yielding $\widehat{\bm{\mathcal{K}}} = (\hat{\kappa}_{qq'})_{1 \le q,q' \le Q}$ that are then used to recover a corresponding latent dependence specification. For each endpoint pair $(q, q')$, the choice of concordance measure depends on whether censoring is present. For uncensored endpoint pairs, including continuous, binary, or count outcomes, the observed concordance $\hat{\kappa}_{qq'}$ may be estimated by Kendall's $\tau_b$~\cite{kendall1938new}. For any pair in which at 
least one endpoint is a censored time-to-event outcome, a practical choice is to transform Harrell's C-index~\cite{harrell1982evaluating} as
\[
\hat{\kappa}_{qq'} = 2\hat{C}_{qq'} - 1,
\] which maps the proportion of concordant evaluable pairs to a $[-1,1]$ scale analogous to Somers' $D$. In both cases, the resulting $\hat{\kappa}_{qq'}$ serves as a pairwise calibration target for the numerical recovery of $\bm{\mathcal{R}}$.

Given the full target matrix $\widehat{\bm{\mathcal{K}}}$, the latent dependence parameter $\bm{\mathcal{R}}$ is calibrated numerically so that super samples generated under the chosen copula distributions reproduce the target concordance within a pre-specified tolerance. For bivariate HEs ($Q=2$), this calibration reduces to a one-dimensional bisection search over a single $\rho \in [-1, 1]$; for higher-dimensional settings, each pairwise element $\rho_{qq'}$ of $\bm{\mathcal{R}}$ may be tuned iteratively while holding the remaining elements fixed, cycling until all pairwise concordances converge to their respective targets.
\end{itemize}

Whether the design begins from Strategy~1 or Strategy~2, substantial uncertainty about the dependence structure may remain at the planning stage either because there is no pilot data or the trial data may differ significantly from the pilot data. A practical safeguard is therefore to incorporate a pre-specified blinded sample size re-estimation at a pre-specified time, for example after approximately $1/3$ or $1/2$ of subjects have been enrolled. At that stage, the dependence among the component endpoints may be re-estimated from blinded pooled data using strategy 2 and the sample size updated accordingly, while preserving blinding to treatment assignment. This provides a practical mechanism for refining the original FORSS-based design when the initial dependence assumptions are uncertain.


\section{Simulation Studies}\label{sec:simulation}
\subsection{Objectives and Simulation Design}
The simulation studies were designed for four specific goals:
\begin{description}[
    leftmargin=0pt,
    labelsep=0.8em,
    itemsep=0.35em,
    topsep=0.35em,
    font=\normalfont\bfseries
]
\item[Goal 1: Accuracy of FORSS.]
We evaluated the performance of the FORSS framework by comparing calculated power, obtained from the power formulas using super-sample-based plug-in estimates, with empirical power from independent Monte Carlo trial replications. Empirical type~I error was evaluated to assess whether the nominal significance level was preserved.

\item[Goal 2: Impact of endpoint dependence.]
We examined how endpoint dependence changes the population-level quantities that determine power, including the overall win and loss probabilities, the contribution of lower-priority endpoints as tie-breakers, and the covariance components in the power formulas. This goal isolates the role of dependence by varying the latent correlation parameter while holding the marginal distributions and marginal treatment effects fixed.

\item[Goal 3: Consequences of independence-based planning.]
We assessed the practical consequences of planning under an independence assumption, when the component endpoints are in fact correlated.

\item[Goal 4: Computational behavior.]
We examined the computational behavior of the adaptive super-sample procedure in Algorithm~\ref{alg:forss}, including the number of adaptive replicates, runtime, and memory usage.
\end{description}

Similar to Barnhart et al.~\cite{barnhart_sample_2025}, we considered four two-endpoint HE scenarios, as summarized in Table \ref{tab:scenario_specification}, but with different parameters to better demonstrate the impact of the dependence on the power in the chosen scenarios.  These scenarios were chosen to cover common combinations of component endpoint types and tie-breaking structures in hierarchical endpoints. These four scenarios are (1) two continuous endpoints with specific thresholds used as baseline benchmarks, (2) a continuous endpoint followed by a binary endpoint, to illustrate a mixed hierarchy where the second endpoint is used as a coarse  tie-breaker endpoint, (3) a time-to-event endpoint followed by a continuous endpoint, where the first component is a clinically important higher-priority endpoint subject to censoring, and (4) a binary endpoint followed by a continuous endpoint to examine another scenario with coarse endpoint on the HEs. Latent dependence in the working joint model was introduced through a Gaussian copula with a correlation parameter $\rho$, where we considered different levels of correlation $\rho \in \{0.0, 0.2, 0.4, 0.6, 0.8\}$. 

For each scenario, we determined the total sample size ($N=m+n$) required to achieve 85\% nominal power at two-sided $\alpha=0.05$ under independence ($\rho=0$) using Algorithm~\ref{alg:forss} with super-sample size $N_{\textup{sp}}=2{,}000$,  with convergence thresholds of $\varepsilon_{\tau}=5\times10^{-4}$ and $\varepsilon_{\xi}=1\times10^{-4}$ along with iteration limits $b_{\textup{min}}=100$ and $b_{\textup{max}}=3,000$,, this led to a range of 250 and 850 for the total number of super samples needed (see Table 11 in Appendix). 
This sample size $N$ was then held fixed across all correlation levels to isolate the impact of endpoint dependence on the achieved power and contrast it with the power we obtained when the sample-size is estimated assuming the independence assumption.

For each combination of HE scenario and selected correlation $\rho$, the  "calculated power" was compared with the empirical power based on 10{,}000 independent Monte-Carlo trial replicates generated under $H_A$. We defined the empirical power as the proportion of replicates in which the corresponding null hypothesis $H_0$ is rejected at a two-sided significance level of $\alpha=0.05$. The empirical type I error was evaluated analogously under $H_0$, i.e., as the proportion of  Monte-Carlo replicates that reject $H_0$ when the data were generated under no treatment effect. To account for fluctuations due to the number of Monte Carlo data repetitions, an empirical type I error will be considered significantly different from the nominal level $\alpha = 0.05,$ if it falls outside of the interval $0.05\pm 1.96\sqrt{0.05\times 0.95/10,000} = [0.0457, 0.0543]$. 

To facilitate the interpretation on the observed-data scale, we also reported the observed concordance summary $\widehat{\kappa}$ corresponding to each value of $\rho$. For the HEs with two continuous endpoints, $\widehat{\kappa}$ was estimated by Kendall's $\tau_b$, whereas for HEs involving a time-to-event endpoint, $\widehat{\kappa}$ was obtained by transforming Harrell's C-index via $\widehat{\kappa} = 2\widehat{C}-1$~\cite{harrell1982evaluating}. Finally, to clarify the mechanism by which endpoint dependence alters the overall win statistics and power, we decomposed the overall win and loss probabilities into marginal and conditional components.

Following the Stage 1 marginal-specification setup of FORSS, each scenario was defined by specifying, for each endpoint in the hierarchy, the endpoint type, the control-arm marginal distribution $F_q^{(C)}$, and a clinically interpretable marginal effect size $\Delta_q$, which together determine the treatment-arm marginal distribution $F_q^{(T)}$. Table~\ref{tab:scenario_specification} summarizes the four scenarios.

\begin{table}[ht!]
\centering
\caption{Stage 1 marginal specifications for the four simulation scenarios}
\label{tab:scenario_specification}
\setlength{\tabcolsep}{10pt}
\begin{tabular}{@{}ccllll@{}}
\toprule
Scenario & Endpoint $q$ & Data Type & $F_q^{(C)}$ & $\Delta_q$ & $F_q^{(T)}$ \\
\midrule
\multirow{2}{*}{S1}
& $D_1$ & Continuous & $N(3,10^2)$ & MD$=1$, $\delta_1=8$ & $N(4,10^2)$ \\
& $D_2$ & Continuous & $N(30,15^2)$ & MD$=6$, $\delta_2=6$ & $N(36,15^2)$ \\[0.4cm]
\multirow{2}{*}{S2}
& $D_1$ & Continuous & $N(4,10^2)$ & MD$=2$, $\delta_1=8$ & $N(6,10^2)$ \\
& $D_2$ & Binary & $\mathrm{Bernoulli}(0.3)$ & RD$=0.10$ & $\mathrm{Bernoulli}(0.4)$ \\[0.4cm]
\multirow{2}{*}{S3}
& $D_1$ & TTE & $\mathrm{Exp}(0.036)$ & HR$=0.67$ & $\mathrm{Exp}(0.024)$ \\
& $D_2$ & Continuous & $N(3,14^2)$ & MD$=3$, $\delta_2=6$ & $N(6,14^2)$ \\[0.4cm]
\multirow{2}{*}{S4}
& $D_1$ & Binary & $\mathrm{Bernoulli}(0.3)$ & RD$=0.10$ & $\mathrm{Bernoulli}(0.4)$ \\
& $D_2$ & Continuous & $N(4,10^2)$ & MD$=2$, $\delta_2=8$ & $N(6,10^2)$ \\
\bottomrule
\end{tabular}

\parbox{0.94\textwidth}{ \vspace{.2cm}\footnotesize MD = mean difference; RD = risk difference; HR = hazard ratio; TTE = time-to-event. For continuous endpoints, $\Delta_q$ was specified as a mean difference together with threshold $\delta_q$ used in the pairwise comparison rule. For the time-to-event endpoint in Scenario~3, $\Delta_q$ was specified as a hazard ratio under exponential survival with a 10-month follow-up.}
\end{table}

\subsection{Simulation Results}
\subsubsection{Goal 1: Accuracy of FORSS}
Across all four scenarios and all latent correlation levels, the FORSS framework showed close agreement between calculated and empirical operating characteristics (Table~\ref{tab:all_scenarios_op_char}). For power, the agreement was consistently strong for WR, NB/DOOR, and WO, indicating that the super-sample-based plug-in estimates were sufficiently accurate for use in the closed-form power formulas. For WR, the maximum absolute difference between calculated and empirical power across all scenarios was attained in Scenario~2 at $\rho=0$ and was less than 0.0115. In addition, the empirical type~I error rates remained close to the nominal 5\% level throughout, with only minor Monte Carlo fluctuation across scenarios, win measures, and correlation levels. Taken together, these results confirm that FORSS provides a reliable formula-based power evaluation while preserving an appropriate type~I error control.

The observed concordance summary $\widehat{\kappa}$ also tracked the latent copula correlation $\rho$ in a broadly monotone manner across all four scenarios, providing an interpretable observed-scale summary of the imposed dependence structure (Table~\ref{tab:all_scenarios_op_char}). For the three scenarios involving only non-survival endpoints (S1, S2, and S4), $\widehat{\kappa}$ increased steadily from approximately 0 at $\rho=0$ to about 0.53--0.59 at $\rho=0.8$. In contrast, for the survival--continuous scenario (S3), the transformed Harrell concordance measure $2\widehat{C}-1$ rose more rapidly, reaching 0.742 at $\rho=0.8$. Thus, although the observed dependence summaries were monotone in $\rho$ across all scenarios, the mapping from latent to observed dependence was scenario-specific and depended on the endpoint types involved.

\begin{table}[ht!]
\centering
\caption{Operating Characteristics Across Four Two-Endpoint HE Scenarios at Fixed Sample Sizes}
\label{tab:all_scenarios_op_char}
\setlength{\tabcolsep}{3pt}
\begin{tabular}{@{}ccccccccccc@{}}
\toprule
\multicolumn{2}{c}{Correlation} & \multicolumn{3}{c}{Empirical Type I Error (\%)} & \multicolumn{3}{c}{Calculated Power (\%)} & \multicolumn{3}{c}{Empirical Power (\%)} \\
\cmidrule(lr){1-2} \cmidrule(lr){3-5} \cmidrule(lr){6-8} \cmidrule(lr){9-11}
latent $\rho$ & obs. $\widehat{\kappa}$ & WR & NB \& DOOR & WO & WR & NB \& DOOR & WO & WR & NB \& DOOR & WO \\
\midrule
\multicolumn{11}{@{}l}{\textbf{S1: Continuous + Continuous ($m=n=274$)}}\\
0.0 & 0.000 & 4.99 & 5.12 & 4.99 & 85.03 & 85.19 & 85.17 & 85.13 & 85.37 & 85.13 \\
0.2 & 0.128 & 5.00 & 5.10 & 5.00 & 83.16 & 83.30 & 83.31 & 83.25 & 83.63 & 83.26 \\
0.4 & 0.262 & 4.97 & 5.11 & 4.97 & 83.04 & 83.20 & 83.20 & 83.05 & 83.37 & 83.06 \\
0.6 & 0.410 & 5.02 & 5.19 & 5.02 & 84.90 & 85.11 & 85.09 & 85.12 & 85.40 & 85.13 \\
0.8 & 0.590 & 5.10 & 5.23 & 5.10 & 90.50 & 90.85 & 90.75 & 90.60 & 90.81 & 90.60 \\
\addlinespace[4pt]
\multicolumn{11}{@{}l}{\textbf{S2: Continuous + Binary ($m=n=269$)}}\\
0.0 & 0.001 & 4.92 & 5.11 & 4.97 & 85.05 & 86.01 & 85.98 & 85.80 & 86.05 & 85.81 \\
0.2 & 0.127 & 4.95 & 5.08 & 4.98 & 81.74 & 82.76 & 82.77 & 82.80 & 83.02 & 82.84 \\
0.4 & 0.256 & 4.79 & 4.96 & 4.82 & 78.47 & 79.56 & 79.60 & 79.58 & 79.87 & 79.59 \\
0.6 & 0.389 & 4.88 & 5.04 & 4.89 & 75.42 & 76.55 & 76.62 & 76.38 & 76.74 & 76.40 \\
0.8 & 0.533 & 4.80 & 4.99 & 4.82 & 72.67 & 73.77 & 73.87 & 73.67 & 73.99 & 73.66 \\
\addlinespace[4pt]
\multicolumn{11}{@{}l}{\textbf{S3: Survival + Continuous ($m=n=239$)}}\\
0.0 & 0.001 & 4.85 & 5.04 & 4.85 & 85.05 & 84.28 & 84.27 & 84.04 & 84.33 & 84.09 \\
0.2 & 0.180 & 4.94 & 5.06 & 4.94 & 81.62 & 80.67 & 80.71 & 80.64 & 81.10 & 80.70 \\
0.4 & 0.362 & 4.94 & 5.11 & 4.95 & 78.27 & 77.14 & 77.24 & 77.18 & 77.62 & 77.21 \\
0.6 & 0.548 & 5.01 & 5.22 & 5.02 & 74.78 & 73.49 & 73.63 & 73.91 & 74.31 & 73.94 \\
0.8 & 0.742 & 5.14 & 5.33 & 5.15 & 71.73 & 70.31 & 70.48 & 70.58 & 71.13 & 70.63 \\
\addlinespace[4pt]
\multicolumn{11}{@{}l}{\textbf{S4: Binary + Continuous ($m=n=239$)}}\\
0.0 & 0.001 & 5.09 & 5.24 & 5.10 & 85.14 & 86.07 & 86.04 & 85.64 & 85.91 & 85.65 \\
0.2 & 0.127 & 5.04 & 5.24 & 5.07 & 81.38 & 82.39 & 82.40 & 81.88 & 82.22 & 81.88 \\
0.4 & 0.256 & 4.96 & 5.17 & 5.00 & 77.47 & 78.53 & 78.60 & 78.21 & 78.63 & 78.25 \\
0.6 & 0.390 & 5.21 & 5.41 & 5.23 & 73.43 & 74.55 & 74.66 & 74.25 & 74.63 & 74.28 \\
0.8 & 0.533 & 5.36 & 5.46 & 5.37 & 69.31 & 70.44 & 70.59 & 70.34 & 70.66 & 70.36 \\
\bottomrule
\end{tabular}
\parbox{0.92\textwidth}{\vspace{.2cm}\footnotesize \textit{Note:} WR: Win Ratio; NB: Net Benefit; DOOR: Desirability of Outcome Ranking; WO: Win Odds. These results are based on $N_{sp}=2000$ runs. Adaptive Monte Carlo estimation used $B_{\min}=100$, $B_{\max}=3,000$, $\varepsilon_{\tau}=5\times 10^{-4}$, and $\varepsilon_{\xi}=10^{-4}$. Observed $\widehat{\kappa}$ is Kendall's tau for S1, S2, and S4; for S3 it was estimated as $2\widehat{C}-1$ from Harrell's C-index. Some S3 runs reached \texttt{B\_MAX\_REACHED}; see the computational diagnostics table for details.}
\end{table}

\subsubsection{Goal 2: Impact of endpoint dependence.}

To understand why power changed across correlation levels, Table~\ref{tab:decomposition} decomposes the overall pairwise comparison into three layers: the marginal win, loss, and tie probabilities on the first endpoint $D_1$; the conditional win, loss, and tie probabilities on the second endpoint given a tie on $D_1$; and the resulting overall win, loss, and tie probabilities ($\tau_w, \tau_l$ and $\tau_\Omega$). As expected, across S1-S4, the marginal probabilities on $D_1$ remained essentially unchanged as $\rho$ increased, indicating that dependence did not materially affect the contribution of the first endpoint itself. Instead, the impact of dependence operated primarily through the second endpoint, conditional on a tie on $D_1$, that is, through how often the lower-priority endpoint $D_2$ could still resolve those tied pairs into wins or losses rather than just additional ties. In particular, as $\rho$ increased, the conditional tie probability on $D_2$ increased consistently, whereas the conditional win and loss probabilities both decreased, with the relative rates of decline differing across scenarios. This relative imbalance is the key mechanism by which dependence either attenuated or amplified the net treatment contrast, and hence the achieved power, under a fixed planned sample size.

\begin{table}[ht!]
\centering
\caption{Decomposition of Win Probabilities Across Four Two-Endpoint HE Scenarios}
\label{tab:decomposition}
\setlength{\tabcolsep}{3pt}
\begin{tabular}{@{}cccccccccccccc@{}}
\toprule
latent $\rho$ & \multicolumn{3}{c}{On $D_1$ (\%)} & \multicolumn{3}{c}{Given Tie on $D_1$ (\%)} & \multicolumn{3}{c}{Overall (\%)} & \multicolumn{4}{c}{Overall Measures} \\
\cmidrule(lr){2-4} \cmidrule(lr){5-7} \cmidrule(lr){8-10} \cmidrule(l){11-14}
 & Win & Loss & Tie & Win & Loss & Tie & Win $\tau_w$ & Loss $\tau_l$ & Tie $\tau_\Omega$ & WR & NB & WO & DOOR \\
\midrule
\multicolumn{14}{@{}l}{\textbf{S1: Continuous + Continuous}}\\
0.0 & 31.02 & 26.23 & 42.75 & 50.02 & 28.56 & 21.41 & 52.41 & 38.44 & 9.15 & 1.363 & 0.140 & 1.325 & 0.570 \\
0.2 & 31.02 & 26.23 & 42.75 & 49.51 & 28.65 & 21.84 & 52.19 & 38.47 & 9.34 & 1.356 & 0.137 & 1.318 & 0.569 \\
0.4 & 31.04 & 26.22 & 42.73 & 48.94 & 27.95 & 23.11 & 51.95 & 38.17 & 9.88 & 1.361 & 0.138 & 1.320 & 0.569 \\
0.6 & 31.02 & 26.25 & 42.73 & 48.20 & 26.05 & 25.75 & 51.62 & 37.38 & 11.00 & 1.381 & 0.142 & 1.332 & 0.571 \\
0.8 & 31.03 & 26.24 & 42.73 & 46.95 & 21.55 & 31.51 & 51.09 & 35.45 & 13.46 & 1.441 & 0.156 & 1.371 & 0.578 \\
\addlinespace[4pt]
\multicolumn{14}{@{}l}{\textbf{S2: Continuous + Binary}}\\
0.0 & 33.56 & 23.97 & 42.47 & 27.98 & 18.02 & 54.00 & 45.44 & 31.62 & 22.93 & 1.437 & 0.138 & 1.321 & 0.569 \\
0.2 & 33.56 & 23.97 & 42.47 & 26.98 & 18.31 & 54.71 & 45.02 & 31.75 & 23.24 & 1.418 & 0.133 & 1.306 & 0.566 \\
0.4 & 33.55 & 23.98 & 42.47 & 25.43 & 17.90 & 56.67 & 44.35 & 31.58 & 24.07 & 1.404 & 0.128 & 1.293 & 0.564 \\
0.6 & 33.55 & 23.99 & 42.46 & 23.06 & 16.59 & 60.35 & 43.34 & 31.03 & 25.63 & 1.397 & 0.123 & 1.281 & 0.562 \\
0.8 & 33.55 & 23.98 & 42.47 & 19.16 & 13.77 & 67.07 & 41.68 & 29.83 & 28.48 & 1.397 & 0.119 & 1.269 & 0.559 \\
\addlinespace[4pt]
\multicolumn{14}{@{}l}{\textbf{S3: Survival + Continuous}}\\
0.0 & 27.08 & 18.05 & 54.87 & 43.96 & 32.49 & 23.55 & 51.20 & 35.88 & 12.92 & 1.427 & 0.153 & 1.362 & 0.577 \\
0.2 & 27.08 & 18.05 & 54.87 & 43.17 & 33.00 & 23.82 & 50.77 & 36.16 & 13.07 & 1.404 & 0.146 & 1.342 & 0.573 \\
0.4 & 27.07 & 18.05 & 54.88 & 42.23 & 33.21 & 24.56 & 50.24 & 36.28 & 13.48 & 1.385 & 0.140 & 1.325 & 0.570 \\
0.6 & 27.07 & 18.06 & 54.87 & 40.98 & 33.09 & 25.93 & 49.56 & 36.21 & 14.23 & 1.369 & 0.133 & 1.308 & 0.567 \\
0.8 & 27.09 & 18.06 & 54.84 & 39.29 & 32.39 & 28.32 & 48.64 & 35.83 & 15.53 & 1.358 & 0.128 & 1.294 & 0.564 \\
\addlinespace[4pt]
\multicolumn{14}{@{}l}{\textbf{S4: Binary + Continuous}}\\
0.0 & 28.00 & 18.00 & 54.00 & 33.54 & 23.99 & 42.47 & 46.11 & 30.96 & 22.93 & 1.489 & 0.151 & 1.357 & 0.576 \\
0.2 & 28.00 & 18.00 & 54.01 & 32.54 & 24.43 & 43.03 & 45.57 & 31.19 & 23.24 & 1.461 & 0.144 & 1.336 & 0.572 \\
0.4 & 27.98 & 18.01 & 54.01 & 31.12 & 24.32 & 44.56 & 44.79 & 31.14 & 24.07 & 1.438 & 0.136 & 1.316 & 0.568 \\
0.6 & 27.98 & 18.01 & 54.01 & 29.00 & 23.55 & 47.45 & 43.64 & 30.73 & 25.63 & 1.420 & 0.129 & 1.297 & 0.565 \\
0.8 & 27.99 & 18.01 & 54.00 & 25.65 & 21.63 & 52.72 & 41.84 & 29.69 & 28.47 & 1.410 & 0.122 & 1.277 & 0.561 \\
\bottomrule
\end{tabular}
\end{table}

\begin{figure}[ht!]
    \centering
    \includegraphics[width=0.6\linewidth]{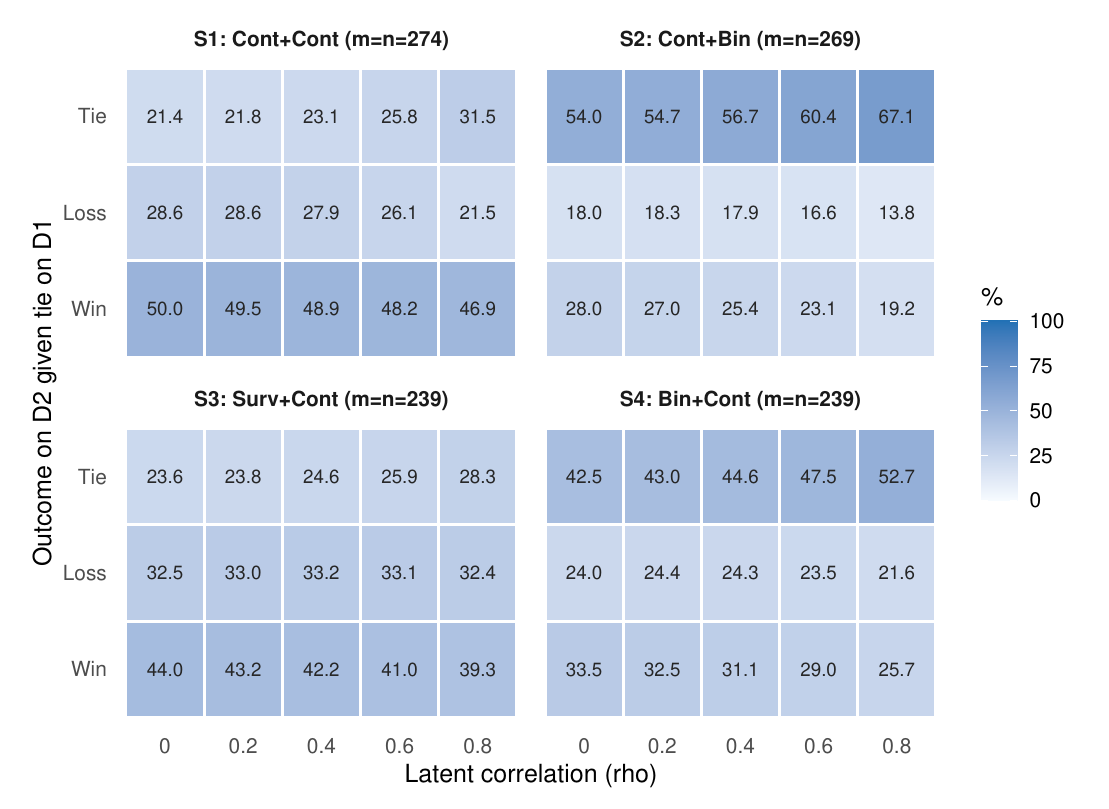}
    \caption{Conditional win, loss, and tie probabilities on $D_2$ among pairs tied on $D_1$, across latent correlation values and two-endpoint HE scenarios.}
    \label{fig:d2_conditional_prob}
\end{figure}

S1 showed a qualitatively different pattern from the other three settings. As $\rho$ increased, the conditional win probability on $D_2$ declined modestly, but the conditional loss probability declined more sharply, while the conditional tie probability increased (Figure~\ref{fig:d2_conditional_prob}). As a result, the overall loss probability $\tau_l$ decreased faster than the overall win probability $\tau_w$, so the net treatment contrast was amplified rather than attenuated. For example, as $\rho$ increased from 0 to 0.8, the overall WR increased from 1.363 to 1.441, accompanied by increases in NB, WO, and DOOR.  Thus, in the S1 setting, stronger dependence did not weaken the overall treatment effect; instead, it reduced overall losses more than overall wins and therefore increased the achieved power. 

By contrast, S2-S4 showed an alternative pattern of attenuation under increasing dependence. In these mixed types of HE settings, the lower-priority endpoint $D_2$ became less effective as a tie-breaker as $\rho$ increased. In S2, for example, the conditional win probability given a tie on $D_1$ decreased from 27.98\% to 19.16\%, whereas the conditional loss probability decreased less dramatically, from 18.02\% to 13.77\%; at the same time, the conditional tie probability increased from 54.00\% to 67.07\%. Similar asymmetric erosion was observed in S3 and S4. In Scenario~3, the decline was particularly concentrated on the conditional win side, whereas the conditional loss probability changed only minimally, making the attenuation especially pronounced. Consequently, the overall treatment effect was progressively compressed: WR declined from 1.437 to 1.397 in S2, from 1.427 to 1.358 in S3, and from 1.489 to 1.410 in S4. Thus, the effect of dependence on power was governed not simply by the presence of correlation itself, but by whether dependence reduced conditional win and loss information symmetrically or asymmetrically, thereby attenuating or amplifying the net treatment contrast contributed by the lower-priority endpoint.

\subsubsection{Goal 3: Consequences of independence-based planning}
\begin{figure}[ht!]
    \centering
    \includegraphics[width=0.6\linewidth]{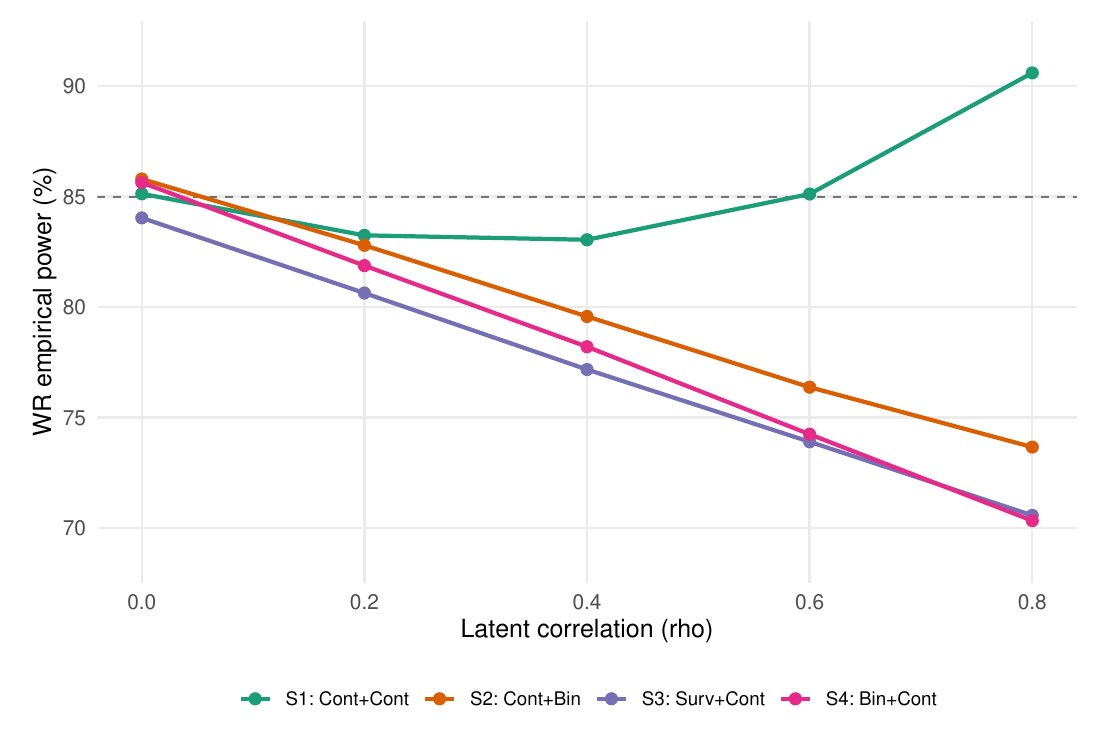}
    \caption{Empirical power of WR across latent correlation levels $\rho$ under independence-based planning. For each scenario, the total sample size $N$ was determined under $\rho=0$ and then held fixed $N$ as $\rho$ increased.}
    \label{fig:wr_emp_power_by_scenario}
\end{figure}

We next examined the practical consequence of following an independence-based sample size planning strategy, as in Barnhart et al.~\cite{barnhart_sample_2025}, when the underlying true endpoint dependence is stronger than assumed at the design stage. Specifically, for each scenario, the total sample size $N$ was first determined under $\rho=0$ to achieve 85\% nominal power. Then, the obtained total sample size $N$ was held fixed as the latent correlation $\rho$ increased (from 0 to 0.8) to estimate the corresponding empirical power. Figure~\ref{fig:wr_emp_power_by_scenario} displays the resulting empirical power of the WR across $\rho$, thereby showing directly how achieved power changes when independence-based planning is applied to correlated hierarchical endpoints.

The practical consequence of independence-based planning was strongly scenario-dependent. In S1, the design planned under independence became mildly conservative at stronger dependence, with empirical power increasing from 85.1\% at $\rho=0$ to 90.6\% at $\rho=0.8$. In contrast, in S2--S4, the same planning strategy became increasingly anti-conservative: empirical power declined from 85.8\% to 73.9\% in S2, from 84.0\% to 70.6\% in S3, and from 85.6\% to 70.3\% in S4 as $\rho$ increased from 0 to 0.8. Thus, ignoring dependence led to mild overpowering in the S1 setting, but substantial underpowering in the remaining settings.

These results also help explain why our conclusions differ from those of Barnhart et al.~\cite{barnhart_sample_2025}, who reported only a modest impact of dependence on power in the settings they examined. Our findings do not contradict theirs; rather, they suggest that the practical importance of dependence is highly scenario-specific. In some settings, dependence may remain relatively unimportant because the lower-priority endpoint contributes little additional tie-breaking information or because dependence affects conditional win and loss probabilities in a broadly symmetric way. In other settings, however, dependence can materially alter power when the lower-priority endpoint remains influential and dependence changes conditional win and loss information asymmetrically.

One possible explanation for the difference between our findings and earlier
formula-based planning results is that existing formulas use simplified
variance approximations. In particular, the Yu--Ganju formula approximates the
variance of \(\log(\widehat{\mathrm{WR}})\) using a tie-probability-based
quantity, and Barnhart et al. use this approximation for WR and extend the same
idea to other win measures through algebraic relationships among the win
measures~\cite{yu_sample_2022, barnhart_sample_2025}. Therefore, the apparent
impact of dependence could, in principle, be confounded with two variance
approximations: replacing the alternative variance quantity by the null variance
quantity, and further replacing the full U-statistic variance quantity by a
tie-only function of the overall tie probability.

To examine this possibility, we conducted the nested sensitivity analysis in
Appendix~\ref{app:equal_variance_sensitivity}. For each fixed value of
\(\rho\), all formula-based calculations used the same plug-in effect size
\(\widehat{\Delta}\), the same sample size, and the same marginal endpoint
specifications. We then changed only the variance quantity used in the power
calculation: the FORSS alternative variance quantity
\(\widehat{\mathcal A}_A\), the corresponding null variance quantity
\(\widehat{\mathcal A}_0\), or the Yu--Ganju/Barnhart tie-only quantity
\(\widehat{\mathcal A}_{YG}\). This nested comparison separates the impact of
the null-variance substitution from the additional tie-only variance
approximation.

The appendix results show that replacing \(\widehat{\mathcal A}_A\) by
\(\widehat{\mathcal A}_0\) changed the calculated power only modestly relative
to the much larger empirical power changes observed as \(\rho\) increased.
Thus, the underpowering observed in S2--S4 is not explained primarily by the
equal-variance approximation. Rather, it reflects the way the dependence
structure changes the population-level plug-in quantities for the hierarchical
endpoint, especially the overall win/loss contrast, with any additional impact
from the tie-only variance approximation being secondary in these scenarios.

In practice, the internal dependence structure of planned HEs is typically not known well enough at the design stage to determine in advance which of these situations applies. The sensitivity analysis above suggests that simply relying
on an independence-based calculation, or on a variance approximation that
depends only on the overall tie probability, may not reveal the full range of
achieved power when dependence changes the population win/loss functionals. We therefore recommend using FORSS with dependence sensitivity analysis or pilot-data-based calibration described in Section~\ref{subsec:dependence-forss}, rather than relying on an independence-based sample size calculation alone. Specifically, Strategy~1 provides a practical approach when only a plausible range of dependence values can be elicited, whereas Strategy~2 can be used when pilot or historical data are available to calibrate the working dependence structure to observed concordance summaries. These strategies help reduce the risk of materially underpowered designs when dependence is more consequential than assumed under independence.

\subsubsection{Goal 4: Computational Behavior of FORSS}

To assess the practical feasibility of FORSS, we summarized the computational behavior of the adaptive super-sample procedure across all four scenarios and all latent correlation levels considered. Table~\ref{tab:all_scenarios_computation_summary} reports the final number of replicates $B_\textup{final}$, runtime, worker-level memory usage, and the final stopping diagnostics $\widehat{\mathrm{SE}}_{\max}^{(\tau)}$. All computations were carried out on FORSS using 18 CPU cores in parallel for each scenario. Overall, these results indicate that the computational burden of FORSS is practically manageable for above scenarios with $N_\textup{sp}=2,000$. For completeness, a matched-threshold comparison between $N_\textup{sp}=8{,}000$ runs and the standardized $N_\textup{sp}=2{,}000$ runs at $\rho=0$ is reported in Appendix~\ref{app:nsp8000_vs_2000}. It shows that the smaller super-population size $N_\textup{sp}$ substantially reduces worker memory usage and is generally faster in practice. 

\begin{table}[ht!]
\centering
\caption{Summary of Adaptive Monte Carlo Computational Diagnostics Across Four Scenarios}
\label{tab:all_scenarios_computation_summary}
\setlength{\tabcolsep}{3pt}
\begin{tabular}{@{}ccccc@{}}
\toprule
Scenario & $b_{\text{final}}$ & Time (min) & Worker max (GB) & Max SE$(\tau)$ \\
\midrule
S1 & 600--950 & 3.3--5.3 & 0.79--0.80 & $3.74\times 10^{-4}$--$4.98\times 10^{-4}$  \\
S2 & 1000--1350 & 5.5--7.4 & 0.78--0.81 & $2.99\times 10^{-4}$--$5.00\times 10^{-4}$ \\
S3 & 850--3,000 & 4.8--43.1 & 0.80--0.92 & $4.93\times 10^{-4}$--$5.14\times 10^{-4}$ \\
S4 & 1050--1350 & 5.7--7.4 & 0.78--0.81 & $3.10\times 10^{-4}$--$5.00\times 10^{-4}$ \\
\bottomrule
\end{tabular}
\parbox{0.92\textwidth}{ \vspace{.2cm}\footnotesize S1: continuous + continuous; S2: continuous + binary; S3: survival + continuous; S4: binary + continuous. Ranges are taken over latent correlations $\rho \in \{0, 0.2, 0.4, 0.6, 0.8\}$. These summaries correspond to the standardized $N_\textup{sp}=2000$ runs. The stopping parameters were $b_{\min}=100$, $b_{\max}=3,000$, $\varepsilon_{\tau}=5\times 10^{-4}$, and $\varepsilon_{\xi}=10^{-4}$.}
\end{table}

\begin{figure}
    \centering
    \includegraphics[width=\textwidth]{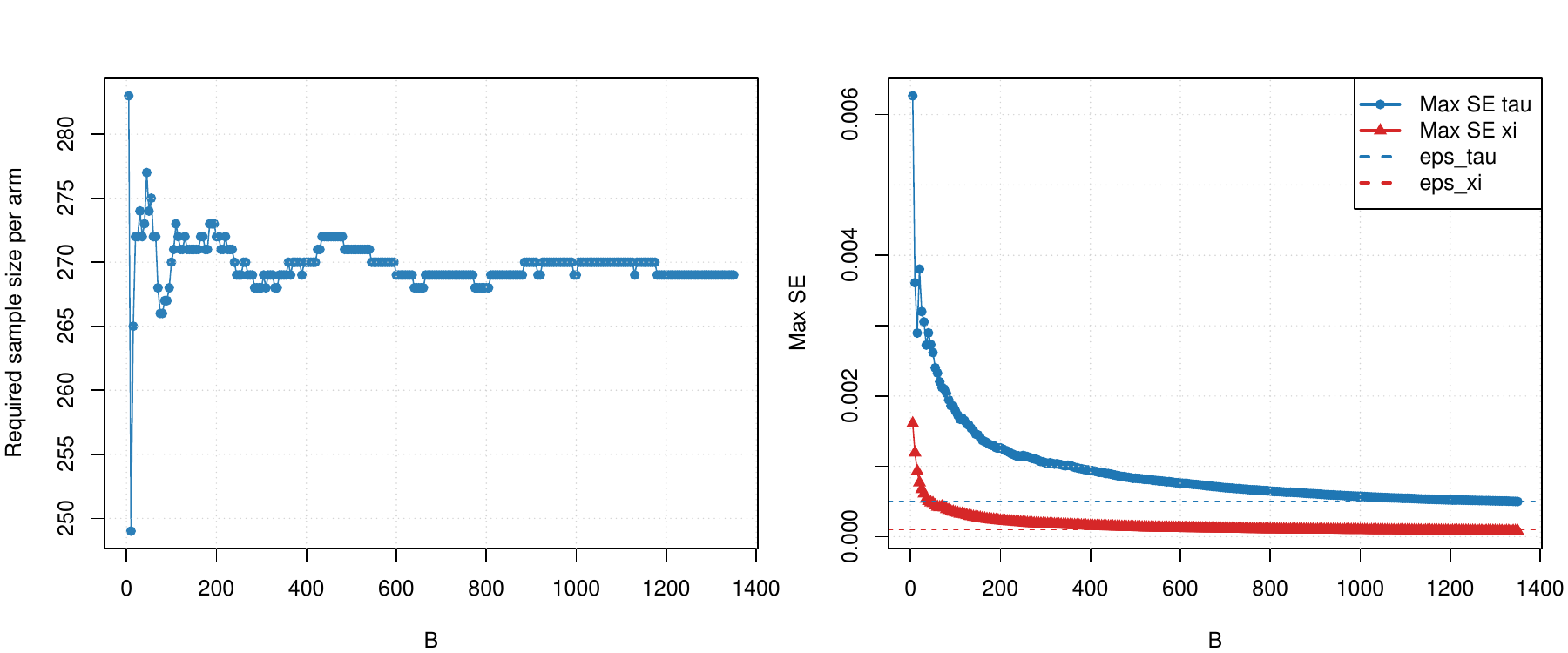}
    \caption{Representative convergence diagnostics for the adaptive super-sample procedure in Scenario 2 (Continuous + Binary) at $\rho=0$. The left panel shows the estimated required sample size per arm across replicates, and the right panel shows the corresponding stopping diagnostics $\widehat{\mathrm{SE}}_{\max}^{(\tau)}$ and $\widehat{\mathrm{SE}}_{\max}^{(\xi)}$ relative to their thresholds.}
    \label{fig:convergence_example}
\end{figure}

Figure~\ref{fig:convergence_example} illustrates a representative convergence profile for Scenario 2 at $\rho=0$. The estimated required sample size stabilizes after an initial transient phase, while the stopping diagnostics decrease monotonically with iteration. 

\section{Data Example: Trial Designs with the HEART-FID Trial as Pilot Data}\label{sec:HEART-FID}

\subsection{Marginal Specification with HEART-FID Data}

We used the HEART-FID trial as pilot data to design a future study with the same three-endpoint hierarchy, rather than to re-analyze the original trial. This setting is especially useful because Barnhart et al.\ previously used HEART-FID to illustrate an independence-based formula approach for sample size and power calculation with win measures. Here, we revisit the same trial setting for a different purpose: to examine how FORSS design calculations behave when the marginal working distributions are held fixed but the working dependence specification is varied. Thus, the HEART-FID example serves as a clinically grounded case study for isolating the role of dependence modeling at the design stage.

The HEART-FID trial randomized ambulatory patients with heart failure and iron deficiency to intravenous ferric carboxymaltose or placebo\cite{mentz_ferric_2023}. Its primary analysis was based on HEs with $Q=3$ endpoints, which consist of time to death within 12 months $D_1$, the number of heart-failure hospitalizations within 12 months $D_2$, and change from baseline to 6 months in 6-minute walk distance (6MWD) $D_3$. This hierarchy places the most severe clinical endpoint first, while allowing lower-priority endpoints to contribute only when higher-priority comparisons are unresolved. Following Barnhart et al.\ \cite{barnhart_sample_2025}, we treated the published HEART-FID summaries based on the first imputed data set as pilot data for a future trial design with the same hierarchical structure. Table~\ref{tab:heartfid_params} summarizes the corresponding endpoint definitions, observed treatment-arm summaries, and the working marginal distributions to be used to illustrate our FORSS approach. Specifically, we assumed $D_1$ follows an exponential distribution, $D_2$ follows a Poisson distribution, and $D_3$ follows a normal distribution. The administrative follow-up horizon for the survival and hospitalization components was fixed at 12 months, and the continuous endpoint was analyzed on its natural scale at 6 months.

\begin{table}[ht]
\centering
\begin{threeparttable}
\caption{Hierarchical structure, observed arm-specific summaries, and working marginal models for the HEART-FID illustration based on the first imputed data set\tnote{a}.}
\label{tab:heartfid_params}
\begin{tabular}{cllcc}
\toprule
 & Outcome & Marginal & Control Arm Input & Treatment Arm Input \\
\midrule
$D_1$ & Time to death within 1 year & Exponential & $p_C = 0.103$ & $p_T = 0.086$\tnote{b} \\
$D_2$\tnote{c} & \# HF hospitalizations within 1 year & Poisson & $\mu_C = 0.332$ & $\mu_T = 0.257$ \\
$D_3$ & $\Delta$ 6MWD from baseline to 6 months\tnote{d}  & Normal & $\mu_C = -24.02, \sigma_C = 101.17$ & $\mu_T = -22.22, \sigma_T = 106.83$ \\
\bottomrule
\end{tabular}
\begin{tablenotes}\footnotesize
    \item[a] All arm-specific summaries in this table are based on the first imputed data set reported by Barnhart et al.\ \cite{barnhart_sample_2025} and are used here as fixed marginal design inputs for the HEART-FID illustration.
    \item[b] The reported values for $D_1$ are 1-year event probabilities. For simulation under an exponential model, we use $\lambda = -\log(1-p)/1$, yielding hazard rates of $\lambda_C = 0.1088$ and $\lambda_T = 0.0899$.
    \item[c] The value for $D_2$ is the observed averaged number of HF hospitalizations adjusted for follow-up time at 1 year.
    \item[d] $\Delta$ 6MWD represents the change from baseline to month 6. Thresholds $\delta_2=\delta_3 = 0$ are assumed for the count and continuous endpoints in deciding wins.
\end{tablenotes}
\end{threeparttable}
\end{table}

\subsection{Working Dependence Specifications}

To examine the role of dependence in study design using pilot data from the HEART-FID trial, we considered three Gaussian-copula working dependence specifications while holding the marginal working distributions in Table~\ref{tab:heartfid_params} fixed. Because the copula is parameterized through latent correlations $\bm{\mathcal R}$, whereas the available HEART-FID dependence summaries are reported on the observed endpoint scale, this illustration also allows us to assess how different observed-to-latent mappings affect the resulting design calculations. For the three-component HE considered here, the Gaussian copula dependence is fully determined by the three off-diagonal latent correlations, which we denote by
$
\bm{\rho}
=
(\rho_{12}, \rho_{13}, \rho_{23}).
$
The corresponding latent correlation matrix $\bm{\mathcal R}$ has unit diagonal entries and these three values as its off-diagonal entries.

First, as a benchmark, we considered an independence specification with latent correlation matrix $\bm{\mathcal R}^{\textup{ind}}=\mathbf I_3$, with $\bm{\rho}^{\textup{ind}}=(0,0,0)$. This setting provides the direct analogue of the independence-based design framework considered by Barnhart et al.\ \cite{barnhart_sample_2025}  and serves as a useful baseline for comparing FORSS approach with existing formula-based approaches. 

Second, we considered a direct observed-input specification, $\bm{\mathcal R}^{\textup{dir}}$. Let $\widehat{\bm{\mathcal K}}^{\textup{pilot}}
= (\hat\kappa_{12}^{\textup{pilot}},
 \hat\kappa_{13}^{\textup{pilot}},
 \hat\kappa_{23}^{\textup{pilot}}) 
= (-0.22, 0.52, -0.10)$ denote the estimated observed-scale pairwise concordance summaries from the HEART-FID pilot data. For endpoint pairs involving the time-to-event component $D_1$ with censoring, we can not use correlation of the observed $D_1$ for dependency specification due to censoring. Instead, we use Harrell's C-index, $\hat C_{1q}$, as the dependence measure,  and rescaled it to the $[-1,1]$ concordance scale as $\hat\kappa_{1q}^{\textup{pilot}}=2\hat C_{1q}-1$ for $q=2,3$; for the non-time-to-event pair $(D_2,D_3)$, we used Kendall's $\tau_b$ as $\hat\kappa_{23}^{\textup{pilot}}$. In the direct observed-input specification, denoted by $\bm{\mathcal R}^{\textup{dir}}$, these observed-scale summaries were used directly as the latent Gaussian copula correlations; that is, $ \bm{\rho}^{\textup{dir}} =\widehat{\bm{\mathcal K}}^{\textup{pilot}}.$ This specification is practical but may be naive, because observed-scale concordance summaries are not, in general, equal to the latent correlations in a Gaussian copula.

Third, we considered an observed-scale calibrated specification,  $\bm{\mathcal R}^{\textup{cal}}$, following the pilot-data-based calibration strategy in Section~\ref{subsec:dependence-forss}. Let 
$\bar{\bm{\mathcal K}}^{\textup{sim}}(\bm{\rho})$ denote the average observed-scale 
association summaries induced by simulated data generated from the Gaussian copula 
working model with latent correlation vector $\bm{\rho}$. The calibrated latent 
correlations were obtained by
\[
\bm{\rho}^{\textup{cal}}
=
\arg\min_{\bm{\rho}\in\mathcal{P}_3}
\left\|
\bar{\bm{\mathcal K}}^{\textup{sim}}(\bm{\rho})
-
\widehat{\bm{\mathcal K}}^{\textup{pilot}}
\right\|_2^2,
\]
where $\mathcal{P}_3$ denotes the set of values of 
$(\rho_{12},\rho_{13},\rho_{23})$ that define a valid $3\times 3$ correlation matrix. Calibration fit was assessed by the maximum absolute discrepancy
\[
d_{\max}
=
\left\|
\bar{\bm{\mathcal K}}^{\textup{sim}}(\bm{\rho}^{\textup{cal}})
-
\widehat{\bm{\mathcal K}}^{\textup{pilot}}
\right\|_{\infty}.
\]
In the HEART-FID illustration, $d_{\max}<0.001$, and the Monte Carlo standard errors 
of the fitted observed-scale association summaries were all below $0.003$. This provides 
a practical route for incorporating observed dependence information from historical or 
pilot data, such as HEART-FID, so that the simulated data have a dependence structure 
similar to that observed in the pilot data.

\subsection{Power Comparisons of Designs based on Different Dependence Structure}

Using the HEART-FID trial as preliminary data, assuming independence among the components of the HE, FORSS approach yielded a fixed sample size of $m=n=1,244$ per arm (total $N=2,488$), matching the total sample size considered by Barnhart et al.~\cite{barnhart_sample_2025}, with a target power of 85\%. With the fixed total sample size, we compute power with FORSS across the three working dependence specifications. All three settings used the same standardized FORSS tuning parameters: $N_{\textup{sp}}=2,000$ per arm, $b_{\min}=100$, $b_{\max}=3,000$, $\varepsilon_{\tau}=1\times 10^{-3}$, and $\varepsilon_{\xi}=10^{-4}$. These settings were chosen to keep Monte Carlo error in the plug-in estimates small while maintaining manageable computation. Table~\ref{tab:heartfid_power_main} displays the calculated power via FORSS and the empirical power via 10,000 simulated data sets for testing $H_0$ of WR=1.

\begin{table}[ht!]
\centering
\caption{Win ratio operating characteristics under three working dependence specifications for the HEART-FID illustration ($m=n=1,244$ per arm).}
\label{tab:heartfid_power_main}
\begin{threeparttable}
\setlength{\tabcolsep}{3.2pt}
\renewcommand{\arraystretch}{1.05}
\begin{tabular}{@{}lcccccc@{}}
\toprule
Specification & Latent $\bm{\rho}$ \tnote{a} & Implied obs. $\bar{\bm{\mathcal K}}^{\textup{sim}}$ & Avg.\ WR & Calc.\ power & Emp.\ power & Type I error \\
\midrule
$\bm{\mathcal R}^{\textup{ind}}$ & $(0.00,\,0.00,\,0.00)$ & $(0.00,\,0.00,\,0.00)$ & 1.149 & 85.00 & 84.43 & 4.85 \\
$\bm{\mathcal R}^{\textup{dir}}$ & $(-0.22, 0.52, -0.10)$ & $(-0.15, 0.55, -0.06)$ & 1.132 & 76.20 & 76.59 & 5.28 \\
$\bm{\mathcal R}^{\textup{cal}}$ & $(-0.30, 0.49, -0.17)$ & $(-0.21, 0.53, -0.10)$ & 1.130 & 74.87 & 75.13 & 5.09 \\
\bottomrule
\end{tabular}
\begin{tablenotes}\footnotesize
\item[a] $\bm{\rho}=(\rho_{12},\rho_{13},\rho_{23})$ denotes the off-diagonal entries of the latent Gaussian copula correlation matrix for endpoint pairs $(D_1,D_2)$, $(D_1,D_3)$, and $(D_2,D_3)$.
\item Note: Avg.\ WR denotes average simulated win ratio. Power and type I error are percentages.
\end{tablenotes}
\end{threeparttable}
\end{table}

Under the independence specification $\bm{\mathcal R}^{\textup{ind}}$, the average simulated WR was 1.149, with calculated and empirical powers of 85.00\% and 84.43\%, respectively. These results are consistent with the target power under the independence-based design by Barnhart et al.\cite{barnhart_sample_2025} Under $\bm{\mathcal R}^{\textup{dir}}$, where the observed-scale concordance summaries from the HEART-FID pilot data were directly used as latent correlations $\mathbf{\rho}$. The implied observed-scale summaries in the simulated data were 
$\bar{\bm{\mathcal K}}^{\textup{sim}}=(-0.15,\,0.55,\,-0.06)$, which did not exactly reproduce the pilot-data target 
$\widehat{\bm{\mathcal K}}^{\textup{pilot}}=(-0.22,\,0.52,\,-0.10)$.  The average simulated WR decreased from 1.149 under independence assumption to 1.132, and both calculated and empirical power decreased to about 76\%. Under $\bm{\mathcal R}^{\textup{cal}}$, the calibrated latent correlations $\bm{\rho}^{\textup{cal}}=(-0.30,\,0.49,\,-0.17)$ produced implied observed-scale summaries 
$\bar{\bm{\mathcal K}}^{\textup{sim}}=(-0.21,\,0.53,\,-0.10)$, which closely matched the pilot-data target. The average simulated WR decreased further to 1.130, with calculated and empirical powers decreasing slightly further to around 75\%. Across all three specifications, the type I error rates remained close to the nominal 5\% level, and the calculated powers closely matched the empirical powers, supporting the accuracy of FORSS under these working dependence structures.

These results suggest that a sample size calculated under independence among HE components may be underpowered when clinically relevant endpoint dependence is incorporated. They also illustrate that using observed-scale summaries directly as latent Gaussian copula correlations may not reproduce the intended observed-scale dependence structure, motivating the calibrated specification.

\section{Discussion}\label{sec:discussion}

The FORSS framework addresses a central challenge of sample size and power calculations in trial design with hierarchical endpoints. To estimate the desired sample size for a trial, we can either run time-consuming simulation studies or use a closed-form formula with short computation but may be inaccurate when key assumptions are mis-specified, because it either relies on unavailable specification of overall win measures and probability of ties, or relies on the independence assumption. The FORSS approach blends the advantages of two approaches together. Methodologically, FORSS complements existing approaches by preserving formula-based approach to reduce computation time while allowing dependence to be explored through a flexible joint working distribution, thus increasing accuracy. Relative to fully simulation-based planning, the super-sample step is used only to estimate the population-level plug-in quantities needed in the sample size and power formulas, rather than to repeatedly simulate complete trials at many candidate sample sizes. Relative to approaches that require direct specification of an overall win measure and probability of ties, FORSS works with marginal inputs that are often easier to elicit from prior literature or subject-matter experts. Super samples via copula allow incorporation of the dependency of the HEs and thus avoid assumption of independence. To facilitate application, we provide a \href{https://github.com/BaoshanZZ}{GitHub} implementation for full local execution. 

A key practical finding of this paper is that the power level can depend materially on how the dependence structure across the components of the HEs is handled at the design stage. In the HEART-FID example, we showed a practical way to use the observed dependence in pilot data to specify the dependence structure for the copula joint distribution used for super samples, while the marginal working distributions for the HEs are held fixed. This approach has an advantage in practice. When only marginal endpoint summaries are available, an independence benchmark remains a useful starting point. However, when prior studies or pilot data also provide observed dependence summaries, it may be important to examine sensitivity to alternative working dependence specifications rather than relying on direct substitution of observed correlations into a latent copula model. In our illustration via HEART-FID example, dependence specification was shown to have impact on the effect size and power.

In this paper, we used a Gaussian copula as the primary illustration, but other copula families or joint working models can be easily implemented when dependence is strongly non-Gaussian or highly non-monotone. Our calibration strategy assumes that historical or pilot-study dependence summaries applies to the future trial population. If that is not the case, an interim blinded sample size re-estimation is recommended by using accumulated blinded data thus far as pilot data. The current framework focuses on fixed-design settings and does not yet incorporate adaptive updating of dependence assumptions during trial conduct, which may require future development.

The present implementation also has limitations related to incomplete observation of outcomes. 
A related limitation is that the present paper focuses on a fixed follow-up setting, in which $S$ serves as a common administrative censoring time, and does not systematically investigate subject-specific censoring or missing data. This choice was made to isolate the role of endpoint dependence in design-stage power calculations. However, the FORSS framework is not restricted to this setting. Because the plug-in quantities used in the sample size and power formulas are estimated from super samples, additional data-generating features can be incorporated at the super-sample generation stage. For example, future extensions could generate subject-specific censoring times or missingness indicators, apply the pre-specified pairwise comparison rules to the resulting observed data, and then estimate the corresponding win/loss probabilities and variance components. Thus, while censoring and missing-data mechanisms are not evaluated empirically in the present paper, they can be accommodated within the same FORSS framework and represent an important direction for future work.

In conclusion, the FORSS framework provides a principled and flexible approach to sample size calculation that accommodates endpoint correlation while working with clinically interpretable inputs for win measures. Our finding that correlation can reduce power by up to 25 percentage points in balanced HE configurations underscores the critical importance of sensitivity analysis. Although the present study focuses on the Gaussian dependence among HE components structure to demonstrate the core mechanism, our accompanying open-source software includes a comprehensive suite of alternative copula specifications, allowing researchers to explore a broader range of correlation assumptions. Future extensions integrating arm sequential designs and adaptive correlation updating could further optimize trial efficiency, but strictly speaking, the current framework and toolkit already offer a robust foundation for adequately powering complex clinical trials with hierarchical endpoints \cite{zhang_sequential_2024}.

\section{Acknowledgment} 

\section*{Data Availability}

The R code used to generate the simulation results and the implementation of the FORSS framework are openly available in the GitHub repository at \url{https://github.com/BaoshanZZ/WinSampleSize}. 

\section*{AI use disclosure}
During manuscript preparation, the authors used ChatGPT (OpenAI) to assist with limited coding support, including drafting and revising analysis/plotting code, and for language editing such as grammar and phrasing improvement. All code, figures, analyses, and manuscript text were reviewed, edited, and verified by the authors, who take full responsibility for the final content.
\bibliography{references}

\newpage
\section{Appendix}

\subsection{Derivation of the variance of win and loss
statistics}\label{sec:Appendix_Var}

In this subsection, we show the details of the derivation of the
variance of win and loss statistics in Formula (3). Starting from the
definition of the win/loss statistic \(U_{\nu}\) in Formula (2), we
obtain

\[\text{var}\left( U_{\nu} \right) = \frac{1}{m^{2}  n^{2}}\text{Cov}\left( \sum_{i = 1}^{m}{\sum_{j = 1}^{n}\varphi_{\nu}^{\text{ij}}},\ \sum_{i^{'} = 1}^{m}{\sum_{j^{'} = 1}^{n}\varphi_{\nu}^{i^{'}j^{'}}} \right) = \frac{1}{m^{2}  n^{2}}\sum_{\substack{i = 1 \\ i^{'} = 1 }}^{m}{\sum_{\substack{j = 1 \\ j^{'} = 1 }}^{n}{\text{cov}\left( \varphi_{\nu}^{\text{ij}},\ \varphi_{\nu}^{i^{'}j^{'}} \right)}}.\]

To simplify the summation, we decompose it into four distinct cases:

\begin{enumerate}
\def\labelenumi{\arabic{enumi})}
\item
  \(i = {\ i}^{'}\ \)and \(j = {\ j}^{'}\):
  \(\displaystyle \sum_{i = 1}^{m}{\sum_{j = 1}^{n}{\text{cov}\left( \varphi_{\nu}^{{ij}},\ \varphi_{\nu}^{{ij}} \right)}} = m  n  \xi_{\nu\nu}^{11}\);
\item
  \(i = {\ i}^{'}\) and \(j \neq {\ j}^{'}\):
  \(\displaystyle \sum_{i = 1}^{m}{\sum_{\substack{j = 1 \\ j^{'} \neq j}}^{n}{\text{cov}\left( \varphi_{\nu}^{{ij}},\ \varphi_{\nu}^{ij^{'}} \right)}} = m  n  \left( n - 1 \right)  \xi_{\nu\nu}^{10}\);
\item
  \(i \neq {\ i}^{'}\) and \(j = {\ j}^{'}\):
  \(\displaystyle \sum_{\substack{i = 1 \\ i^{'} \neq i}}^{m}{\sum_{j = 1}^{n}{\text{cov}\left( \varphi_{\nu}^{{ij}},\ \varphi_{\nu}^{i^{'}j} \right)}} = m  \left( m - 1 \right)  n  \xi_{{\nu\nu}}^{01}\);
\item
  \(i \neq {\ i}^{'}\) and \(j \neq {\ j}^{'}\):
  \(\displaystyle \sum_{\substack{i = 1 \\ i^{'} \neq i}}^{m}{\sum_{\substack{j = 1 \\ j^{'} \neq j}}^{n}{{cov}\left( \varphi_{\nu}^{{ij}},\ \varphi_{\nu}^{i^{'}j^{'}} \right)}} = m  \left( m - 1 \right)  n  \left( n - 1 \right)  0 = 0\).
\end{enumerate}

Combining these four cases yields Formula (3).

\subsection{Variance Estimator for Win Measures}\label{seq:var-estimator}

Existing variance formulas, such as those of Dong et al.,\cite{dong_win_2023} are designed for the analysis stage and depend implicitly on realized data quantities (e.g., the observed proportions of wins, losses, and ties). However, in prospective design stage, these quantities are not directly available and cannot be easily derived from preliminary data or the literature, because win measures are defined through the hierarchical comparison of multiple components within HEs. We therefore propose the following consistent empirical estimators for the variance components required in subsequent power and sample size calculations: 
\begin{align}
    \widehat{\xi}_{uv}^{10} = \frac{1}{mn(n - 1)} \sum_{i = 1}^{m} \sum_{\substack{j_1=1, \\ j_2=1,\\j_1 \neq j_2}}^n \varphi_{uv}^{i;j_1,j_2} - \widehat{\tau}_u \widehat{\tau}_v; ~~~~ 
    \widehat{\xi}_{uv}^{01} = \frac{1}{m(m - 1)n} \sum_{\substack{i_1=1, \\ i_2=1,\\i_1 \neq i_2}}^m \sum_{j = 1}^{n} \varphi_{uv}^{i_1,i_2;j} -  \widehat{\tau}_u \widehat{\tau}_v; ~~  
    \text{ and } ~~~ \widehat{\xi}_{uv}^{11} = \frac{1}{mn} \sum_{i = 1}^{m} \sum_{j = 1}^{n} \varphi_{uv}^{i;j} -  \widehat{\tau}_u \widehat{\tau}_v 
\end{align}
where we define the symmetrized kernel functions as
\[
\varphi_{uv}^{i_1,i_2;j} = \frac{1}{2} \left[ \varphi_{u}^{i_1j} \varphi_{v}^{i_2j} + \varphi_{u}^{i_2j} \varphi_{v}^{i_1j} \right] \quad, \varphi_{uv}^{i;j_1, j_2} = \frac{1}{2} \left[ \varphi_{u}^{ij_1} \varphi_{v}^{ij_2} + \varphi_{u}^{ij_2} \varphi_{v}^{ij_1} \right] \quad\text{and} \quad \varphi_{uv}^{i;j} = \varphi_{u}^{ij} \varphi_{v}^{ij}.
\]

With these consistent estimators, we obtain consistent estimators for the variances $\mathcal{V}(U_{w})$ and $\mathcal{V}(U_{l})$ by substituting
\(\widehat{\xi_{{uv}}^{10}},\ \widehat{\xi_{{uv}}^{01}},\) and
\(\widehat{\xi_{{uv}}^{11}}\) into formulas~\eqref{eq:variance_u}, and \eqref{eq:cov_u}.
These estimators can be applied both for inference on realized data and for prospective design-stage evaluation. Importantly, they can be directly integrated into the FORSS and super sample framework to evaluate the variance of overall win measures under both the null hypothesis $H_0$ and any specified alternative hypothesis $H_A$.

\subsection{Power Calculation formula for NB, WO and
DOOR}\label{Sec:Appendix_PowerCalc}

For the WR, we have
\begin{align*}
1 - \beta &= \Pr\left\lbrack \left| \frac{\log\left( \widehat{\text{WR}} \right)}{\ \sqrt{\ \widehat{\mathcal{V}}_{0}\left( \log\left( \text{WR} \right) \right)}} \right| > z_{1 - {\alpha}/{2}} \middle| H_{A} \right\rbrack\\
&= \ \Pr\left\lbrack \frac{\log\left( \widehat{\text{WR}} \right)}{\ \sqrt{\ \widehat{\mathcal{V}}_{0}\left( \log\left( \text{WR} \right) \right)}} < - z_{1 - {\alpha}/{2}} \middle| H_{A} \right\rbrack + Pr\left\lbrack \frac{\log\left( \widehat{\text{WR}} \right)}{\ \sqrt{\ \widehat{\mathcal{V}}_{0}\left( \log\left( \text{WR} \right) \right)}} > z_{1 - {\alpha}/{2}} \middle| H_{A} \right\rbrack\\
&= \Pr\left\lbrack \frac{\log\left( \widehat{\text{WR}} \right) - \log\left( \displaystyle \frac{\tau_{w}}{\tau_{l}} \right)}{\ \sqrt{\ \widehat{\mathcal{V}}_{A}\left( \log\left( \text{WR} \right) \right)}} < \frac{- z_{1 - {\alpha}/{2}}*\ \sqrt{\ \widehat{\mathcal{V}}_{0}\left( \log\left( \text{WR} \right) \right)} - \log\left( \displaystyle\frac{\tau_{w}}{\tau_{l}} \right)}{\ \sqrt{\ \widehat{\mathcal{V}}_{A}\left( \log\left( \text{WR} \right) \right)}} \middle| H_{A} \right\rbrack\\
&~+ \Pr\left\lbrack \frac{\log\left( \widehat{\text{WR}} \right) - \log\left( \displaystyle\frac{\tau_{w}}{\tau_{l}} \right)}{\ \sqrt{\ \widehat{\mathcal{V}}_{A}\left( \log\left( \text{WR} \right) \right)}} > \frac{z_{1 - {\alpha}/{2}}*\ \sqrt{\ \widehat{\mathcal{V}}_{0}\left( \log\left( \text{WR} \right) \right)} - \log\left( \displaystyle\frac{\tau_{w}}{\tau_{l}} \right)}{\ \sqrt{\ \widehat{\mathcal{V}}_{A}\left( \log\left( \text{WR} \right) \right)}} \middle| H_{A} \right\rbrack\\
&= \Phi\left\lbrack \frac{- z_{1 - {\alpha}/{2}}*\ \sqrt{\ \widehat{\mathcal{V}}_{0}\left( \log\left( \text{WR} \right) \right)} - \log\left( \displaystyle\frac{\tau_{w}}{\tau_{l}} \right)}{\ \sqrt{\ \widehat{\mathcal{V}}_{A}\left( \log\left( \text{WR} \right) \right)}} \right\rbrack + 1 - \Phi\left\lbrack \frac{z_{1 - {\alpha}/{2}}*\ \sqrt{\ \widehat{\mathcal{V}}_{0}\left( \log\left( \text{WR} \right) \right)} - \log\left(\displaystyle \frac{\tau_{w}}{\tau_{l}} \right)}{\ \sqrt{\ \widehat{\mathcal{V}}_{A}\left( \log\left( \text{WR} \right) \right)}} \right\rbrack\\
&= \Phi\left\lbrack \frac{- z_{1 - {\alpha}/{2}}*\ \sqrt{\ \widehat{\mathcal{V}}_{0}\left( \log\left( \text{WR} \right) \right)} - \log\left( \displaystyle\frac{\tau_{w}}{\tau_{l}} \right)}{\ \sqrt{\ \widehat{\mathcal{V}}_{A}\left( \log\left( \text{WR} \right) \right)}} \right\rbrack + \Phi\left\lbrack \frac{- z_{1 - {\alpha}/{2}}*\ \sqrt{\ \widehat{\mathcal{V}}_{0}\left( \log\left( \text{WR} \right) \right)} + \log\left(\displaystyle \frac{\tau_{w}}{\tau_{l}} \right)}{\ \sqrt{\ \widehat{\mathcal{V}}_{A}\left( \log\left( \text{WR} \right) \right)}} \right\rbrack\\
&~\text{ using the fact that } ~~
1 - \Phi\left( x \right) = \ \Phi\left( - x \right). 
\end{align*}

Therefore, the power for the WRn is 

\begin{align*}
1 - \beta \approx \Phi\left\lbrack \frac{- z_{1 - {\alpha}/{2}}*\ \sqrt{\ \widehat{\mathcal{V}}_{0}\left( \log\left( \text{WR} \right) \right)} + \left| \log\left( \displaystyle \frac{\tau_{w}}{\tau_{l}} \right) \right|}{\ \sqrt{\ \widehat{\mathcal{V}}_{A}\left( \log\left( \text{WR} \right) \right)}} \right\rbrack
\end{align*}
\vspace{0.7cm}

Similarly, for the NB, we can show that

\begin{align*}
1 - \beta &= \Phi\left\lbrack \frac{- z_{1 - {\alpha}/{2}}*\ \sqrt{\ \widehat{\mathcal{V}}_{0}\left( \text{NB} \right)} - \left( \tau_{w} - \tau_{l} \right)}{\ \sqrt{\ \widehat{\mathcal{V}}_{A}\left( \text{NB} \right)}} \right\rbrack + \Phi\left\lbrack \frac{- z_{1 - {\alpha}/{2}}*\ \sqrt{\ \widehat{\mathcal{V}}_{0}\left( \text{NB} \right)} + \left( \tau_{w} - \tau_{l} \right)}{\ \sqrt{\ \widehat{\mathcal{V}}_{A}\left( \text{NB} \right)}} \right\rbrack\\
&\approx \Phi\left\lbrack \frac{- z_{1 - {\alpha}/{2}}*\ \sqrt{\ \widehat{\mathcal{V}}_{0}\left( \text{NB} \right)} + \left| \tau_{w} - \tau_{l} \right|}{\ \sqrt{\ \widehat{\mathcal{V}}_{A}\left( \text{NB} \right)}} \right\rbrack
\end{align*}

Following the same approach,  we have for the win odds,

\begin{align*}
1 - \beta &= \Phi\left\lbrack \frac{- z_{1 - {\alpha}/{2}}*\ \sqrt{\ \widehat{\mathcal{V}}_{0}\left( \log\left( \text{WO} \right) \right)} - \log\left( \displaystyle\frac{\tau_{w} + 0.5\tau_{\Omega}}{\tau_{l} + 0.5\tau_{\Omega}} \right)}{\ \sqrt{\ \widehat{\mathcal{V}}_{A}\left( \log\left( \text{WO} \right) \right)}} \right\rbrack+ \ \Phi\left\lbrack \frac{- z_{1 - {\alpha}/{2}}*\ \sqrt{\ \widehat{\mathcal{V}}_{0}\left( \log\left( \text{WO} \right) \right)} + \log\left( \displaystyle\frac{\tau_{w} + 0.5\tau_{\Omega}}{\tau_{l} + 0.5\tau_{\Omega}} \right)}{\ \sqrt{\ \widehat{\mathcal{V}}_{A}\left( \log\left( \text{WO} \right) \right)}} \right\rbrack\\
&\approx \Phi\left\lbrack \frac{- z_{1 - {\alpha}/{2}}*\ \sqrt{\ \widehat{\mathcal{V}}_{0}\left( \log\left( \text{WO} \right) \right)} + \left| \log\left( \displaystyle\frac{\tau_{w} + 0.5\tau_{\Omega}}{\tau_{l} + 0.5\tau_{\Omega}} \right) \right|}{\ \sqrt{\ \widehat{\mathcal{V}}_{A}\left( \log\left( \text{WO} \right) \right)}} \right\rbrack
\end{align*}

Finally, for the
\(\text{DOOR} = \tau_{w} + 0.5\tau_{\Omega} = 0.5\left( 1 + \tau_{w} - \tau_{l} \right)\),
using the fact that
\(\tau_{\Omega} = 1 - \left( \tau_{w} + \tau_{l} \right)\)

\begin{align*}
1 - \beta &= \Pr\left\lbrack \left| \frac{\widehat{\text{DOOR}} - 0.5}{\ \sqrt{\ \widehat{\mathcal{V}}_{0}\left( \text{DOOR} \right)}} \right| > z_{1 - {\alpha}/{2}} \middle| H_{A} \right\rbrack\\
& = \ \Pr\left\lbrack \frac{\widehat{\text{DOOR}} - 0.5}{\ \sqrt{\ \widehat{\mathcal{V}}_{0}\left( \text{DOOR} \right)}} < - z_{1 - {\alpha}/{2}} \middle| H_{A} \right\rbrack + Pr\left\lbrack \frac{\widehat{\text{DOOR}} - 0.5}{\ \sqrt{\ \widehat{\mathcal{V}}_{0}\left( \text{DOOR} \right)}} > z_{1 - {\alpha}/{2}} \middle| H_{A} \right\rbrack\\
& = \Pr\left\lbrack \frac{\widehat{\text{DOOR}} - \left( \tau_{w} + 0.5\tau_{\Omega} \right)}{\ \sqrt{\ \widehat{\mathcal{V}}_{A}\left( \log\left( \text{WR} \right) \right)}} < \frac{- z_{1 - {\alpha}/{2}}*\ \sqrt{\ \widehat{\mathcal{V}}_{0}\left( \text{DOOR} \right)} + 0.5 - {(\tau}_{w} + 0.5\tau_{\Omega})}{\ \sqrt{\ \widehat{\mathcal{V}}_{A}\left( \text{DOOR} \right)}} \middle| H_{A} \right\rbrack\\
&+ \Pr\left\lbrack \frac{\widehat{\text{DOOR}} - \left( \tau_{w} + 0.5\tau_{\Omega} \right)}{\ \sqrt{\ \widehat{\mathcal{V}}_{A}\left( \log\left( \text{WR} \right) \right)}} > \frac{z_{1 - {\alpha}/{2}}*\ \sqrt{\ \widehat{\mathcal{V}}_{0}\left( \text{DOOR} \right)} + 0.5 - {(\tau}_{w} + 0.5\tau_{\Omega})}{\ \sqrt{\ \widehat{\mathcal{V}}_{A}\left( \text{DOOR} \right)}} \middle| H_{A} \right\rbrack\\
&= \Pr\left\lbrack \frac{\widehat{\text{DOOR}} - \left( \tau_{w} + 0.5\tau_{\Omega} \right)}{\ \sqrt{\ \widehat{\mathcal{V}}_{A}\left( \log\left( \text{WR} \right) \right)}} < \frac{- z_{1 - {\alpha}/{2}}*\ \sqrt{\ \widehat{\mathcal{V}}_{0}\left( \text{DOOR} \right)} - 0.5{(\tau}_{w} - \tau_{l})}{\ \sqrt{\ \widehat{\mathcal{V}}_{A}\left( \text{DOOR} \right)}} \middle| H_{A} \right\rbrack\\
&+ 1 - \Pr\left\lbrack \frac{\widehat{\text{DOOR}} - \left( \tau_{w} + 0.5\tau_{\Omega} \right)}{\ \sqrt{\ \widehat{\mathcal{V}}_{A}\left( \log\left( \text{WR} \right) \right)}} < \frac{z_{1 - {\alpha}/{2}}*\ \sqrt{\ \widehat{\mathcal{V}}_{0}\left( \text{DOOR} \right)} - 0.5{(\tau}_{w} - \tau_{l})}{\ \sqrt{\ \widehat{\mathcal{V}}_{A}\left( \text{DOOR} \right)}} \middle| H_{A} \right\rbrack\\
&= \Phi\left\lbrack \frac{- z_{1 - {\alpha}/{2}}*\ \sqrt{\ \widehat{\mathcal{V}}_{0}\left( \text{DOOR} \right)} - 0.5{(\tau}_{w} - \tau_{l})}{\ \sqrt{\ \widehat{\mathcal{V}}_{A}\left( \text{DOOR} \right)}} \right\rbrack + \Phi\left\lbrack \frac{- z_{1 - {\alpha}/{2}}*\ \sqrt{\ \widehat{\mathcal{V}}_{0}\left( \text{DOOR} \right)} + 0.5{(\tau}_{w} - \tau_{l})}{\ \sqrt{\ \widehat{\mathcal{V}}_{A}\left( \text{DOOR} \right)}} \right\rbrack\\
&\approx \Phi\left\lbrack \frac{- z_{1 - {\alpha}/{2}}*\ \sqrt{\ \widehat{\mathcal{V}}_{0}\left( \text{DOOR} \right)} + 0.5{|\tau}_{w} - \tau_{l}|}{\ \sqrt{\ \widehat{\mathcal{V}}_{A}\left( \text{DOOR} \right)}} \right\rbrack = \Phi\left\lbrack \frac{- z_{1 - {\alpha}/{2}}*\ \sqrt{\ \widehat{\mathcal{V}}_{0}\left( \text{NB} \right)} + {|\tau}_{w} - \tau_{l}|}{\ \sqrt{\ \widehat{\mathcal{V}}_{A}\left( \text{NB} \right)}} \right\rbrack
\end{align*}

We have the exact same power formula as the NB.

\subsection{Sensitivity analysis for the null-variance and tie-only variance approximations}
\label{app:equal_variance_sensitivity}

This sensitivity analysis is motivated by the structure of existing formula-based
planning approaches. The Yu--Ganju formula for the log win ratio uses a
tie-probability-based variance approximation that depends on the overall
probability of ties rather than on the full U-statistic covariance structure of
the win and loss indicators.\cite{yu_sample_2022} Barnhart et al. use this
Yu--Ganju approximation for WR and then obtain corresponding formulas for
WO, NB, and DOOR through algebraic relationships among the win
measures.\cite{barnhart_sample_2025,dong_win_2023} Therefore, two distinct
approximations may be involved: first, replacing the alternative variance
quantity by the null variance quantity, and second, replacing the full FORSS
variance quantity by a tie-only quantity depending only on
\(\tau_\Omega\). The goal of this appendix is to separate these two steps.

For each main simulation scenario, we fixed the component-level marginal
distributions and marginal treatment effects and varied only the Gaussian
copula latent correlation, $\rho\in\{0,0.2,0.4,0.6,0.8\}.$ Within each scenario, the sample size was fixed at the value selected under \(\rho=0\) to achieve approximately 85\% power for the win ratio. This diagnostic isolates whether the power changes observed as \(\rho\) increases can be explained by variance approximations alone, rather than by
dependence-induced changes in the population win/loss functionals.

The FORSS two-variance power approximation as used in our main paper is
\begin{align*}
\widehat{\mathrm{Power}}_{A}(\rho)
&=
\Phi \left(
\frac{
-z_{1-\alpha/2}\sqrt{\widehat{\mathcal A}_{0}(\rho)}
+
\sqrt{m}\,\widehat{\Delta}(\rho)
}{
\sqrt{\widehat{\mathcal A}_{A}(\rho)}
}
\right).
\end{align*}

To isolate the null-variance approximation, we replace the alternative
large-sample variance quantity by the null quantity:
\begin{align*}
\widehat{\mathrm{Power}}_{0}(\rho)
&=
\Phi \left(
-z_{1-\alpha/2}
+
\frac{\sqrt{m}\,\widehat{\Delta}(\rho)}
{\sqrt{\widehat{\mathcal A}_{0}(\rho)}}
\right).
\end{align*}

For the Yu--Ganju/Barnhart diagnostic,\cite{yu_sample_2022,barnhart_sample_2025}
we further replace the FORSS null variance quantity by the tie-only
large-sample variance quantity \(\widehat{\mathcal A}_{YG}(\rho)\):
\begin{align*}
\widehat{\mathrm{Power}}_{YG}(\rho)
&=
\Phi \left(
-z_{1-\alpha/2}
+
\frac{\sqrt{m}\,\widehat{\Delta}(\rho)}
{\sqrt{\widehat{\mathcal A}_{YG}(\rho)}}
\right).
\end{align*}
For compactness, we denote these three calculated powers by
\[
P_A(\rho)=\widehat{\mathrm{Power}}_A(\rho),\qquad
P_0(\rho)=\widehat{\mathrm{Power}}_0(\rho),\qquad
P_{YG}(\rho)=\widehat{\mathrm{Power}}_{YG}(\rho).
\]
Since \(N=(1+r)m\), where \(r=n/m\), the corresponding tie-only large-sample variance
quantities on the \(\mathcal A\)-scale are as follows. For \(\log(\mathrm{WR})\),
\[
\widehat{\mathcal A}_{\mathrm{WR},YG}(\rho)
=
\frac{
4\{1+\widehat{\tau}_{\Omega,A}(\rho)\}(1+r)
}{
3r\{1-\widehat{\tau}_{\Omega,A}(\rho)\}
}.
\]
For NB, the corresponding tie-only large-sample variance quantity is
\[
\widehat{\mathcal A}_{\mathrm{NB},YG}(\rho)
=
\frac{
\{1+\widehat{\tau}_{\Omega,A}(\rho)\}
\{1-\widehat{\tau}_{\Omega,A}(\rho)\}(1+r)
}{
3r
}.
\]

Thus, the three formula-based power calculations form a nested diagnostic.
The standard FORSS calculation \(P_A\) uses the null variance quantity for the
rejection boundary and the alternative variance quantity under the alternative.
The FORSS-\(H_0\) calculation \(P_0\) uses the same plug-in effect size
\(\widehat{\Delta}(\rho)\), but replaces the alternative variance quantity by
\(\widehat{\mathcal A}_0(\rho)\). Therefore, \(P_0-P_A\) isolates the impact
of the null-variance substitution. The Yu--Ganju/Barnhart calculation
\(P_{YG}\) further replaces \(\widehat{\mathcal A}_0(\rho)\) by the tie-only
quantity \(\widehat{\mathcal A}_{YG}(\rho)\). Accordingly,
\(\widehat{\mathcal A}_0/\widehat{\mathcal A}_A\) summarizes the
null-to-alternative variance comparison, whereas
\(\widehat{\mathcal A}_{YG}/\widehat{\mathcal A}_0\) summarizes the additional
tie-only variance compression.

\begin{table}[]
\centering
\caption{Sensitivity analysis separating the null-variance substitution from
the Yu--Ganju/Barnhart tie-only variance approximation across the four main
simulation scenarios.}
\label{tab:appendix_equal_variance_sensitivity_all}
\setlength{\tabcolsep}{5pt}
\footnotesize
\begin{tabular}{@{}ccccccccccc@{}}
\toprule
Win Measure & Lat. $\rho$ & $\widehat{\Delta}$ & $\widehat{\tau}_{\Omega,A}$ & $\widehat{\mathcal A}_0/\widehat{\mathcal A}_A$ & $\widehat{\mathcal A}_{YG}/\widehat{\mathcal A}_0$ & Emp. Power & $P_A$ & $P_0$ & $P_{YG}$ & $P_0-P_A$ \\
\midrule
\addlinespace[3pt]
\multicolumn{11}{@{}l}{\textbf{S1: Continuous + Continuous ($m=n=274$)}}\\
WR & 0.0 & 0.310 & 0.092 & 0.989 & 1.097 & 85.1 & 85.0 & 85.2 & 81.8 & 0.1 \\
WR & 0.2 & 0.305 & 0.093 & 0.990 & 1.081 & 83.2 & 83.2 & 83.3 & 80.4 & 0.1 \\
WR & 0.4 & 0.308 & 0.099 & 0.991 & 1.064 & 83.0 & 83.0 & 83.2 & 80.8 & 0.1 \\
WR & 0.6 & 0.323 & 0.110 & 0.991 & 1.047 & 85.1 & 84.9 & 85.0 & 83.4 & 0.1 \\
WR & 0.8 & 0.366 & 0.135 & 0.994 & 1.024 & 90.6 & 90.5 & 90.6 & 89.9 & 0.1 \\ \hdashline
NB & 0.0 & 0.140 & 0.092 & 1.024 & 1.107 & 85.4 & 85.2 & 84.9 & 81.2 & -0.3 \\
NB & 0.2 & 0.137 & 0.093 & 1.024 & 1.090 & 83.6 & 83.3 & 83.0 & 79.7 & -0.3 \\
NB & 0.4 & 0.138 & 0.099 & 1.024 & 1.074 & 83.4 & 83.2 & 82.9 & 80.2 & -0.3 \\
NB & 0.6 & 0.142 & 0.110 & 1.026 & 1.059 & 85.4 & 85.1 & 84.8 & 82.7 & -0.3 \\
NB & 0.8 & 0.156 & 0.135 & 1.030 & 1.045 & 90.8 & 90.8 & 90.5 & 89.3 & -0.3 \\
\addlinespace[2pt]
\multicolumn{11}{@{}l}{\textbf{S2: Continuous + Binary ($m=n=269$)}}\\
WR & 0.0 & 0.363 & 0.229 & 1.012 & 1.077 & 85.8 & 85.0 & 84.9 & 82.2 & -0.1 \\
WR & 0.2 & 0.349 & 0.232 & 1.012 & 1.067 & 82.8 & 81.7 & 81.6 & 79.1 & -0.1 \\
WR & 0.4 & 0.340 & 0.241 & 1.012 & 1.057 & 79.6 & 78.5 & 78.3 & 76.1 & -0.1 \\
WR & 0.6 & 0.334 & 0.256 & 1.011 & 1.049 & 76.4 & 75.4 & 75.3 & 73.3 & -0.1 \\
WR & 0.8 & 0.335 & 0.285 & 1.007 & 1.043 & 73.7 & 72.7 & 72.6 & 70.8 & -0.1 \\
\hdashline
NB & 0.0 & 0.138 & 0.229 & 1.012 & 1.132 & 86.1 & 86.0 & 85.9 & 81.4 & -0.1 \\
NB & 0.2 & 0.133 & 0.232 & 1.009 & 1.119 & 83.0 & 82.8 & 82.7 & 78.3 & -0.1 \\
NB & 0.4 & 0.128 & 0.241 & 1.006 & 1.109 & 79.9 & 79.6 & 79.5 & 75.3 & -0.1 \\
NB & 0.6 & 0.123 & 0.256 & 1.003 & 1.100 & 76.7 & 76.5 & 76.5 & 72.5 & -0.0 \\
NB & 0.8 & 0.119 & 0.285 & 0.999 & 1.093 & 74.0 & 73.8 & 73.8 & 70.0 & 0.0 \\
\addlinespace[2pt]
\multicolumn{11}{@{}l}{\textbf{S3: Survival + Continuous ($m=n=239$)}}\\
WR & 0.0 & 0.356 & 0.129 & 0.951 & 1.047 & 84.0 & 85.1 & 85.7 & 84.0 & 0.6 \\
WR & 0.2 & 0.339 & 0.131 & 0.953 & 1.048 & 80.6 & 81.6 & 82.2 & 80.4 & 0.6 \\
WR & 0.4 & 0.326 & 0.135 & 0.955 & 1.051 & 77.2 & 78.3 & 78.8 & 76.8 & 0.5 \\
WR & 0.6 & 0.314 & 0.142 & 0.956 & 1.054 & 73.9 & 74.8 & 75.3 & 73.1 & 0.5 \\
WR & 0.8 & 0.306 & 0.155 & 0.957 & 1.059 & 70.6 & 71.7 & 72.2 & 69.7 & 0.4 \\
\hdashline
NB & 0.0 & 0.153 & 0.129 & 1.037 & 1.016 & 84.3 & 84.3 & 83.8 & 83.3 & -0.4 \\
NB & 0.2 & 0.146 & 0.131 & 1.034 & 1.016 & 81.1 & 80.7 & 80.3 & 79.6 & -0.4 \\
NB & 0.4 & 0.140 & 0.135 & 1.032 & 1.017 & 77.6 & 77.1 & 76.8 & 76.1 & -0.4 \\
NB & 0.6 & 0.133 & 0.142 & 1.031 & 1.019 & 74.3 & 73.5 & 73.2 & 72.4 & -0.3 \\
NB & 0.8 & 0.128 & 0.155 & 1.030 & 1.024 & 71.1 & 70.3 & 70.0 & 69.0 & -0.3 \\
\addlinespace[2pt]
\multicolumn{11}{@{}l}{\textbf{S4: Binary + Continuous ($m=n=239$)}}\\
WR & 0.0 & 0.398 & 0.229 & 1.004 & 1.010 & 85.6 & 85.1 & 85.1 & 84.7 & -0.0 \\
WR & 0.2 & 0.379 & 0.232 & 1.005 & 1.012 & 81.9 & 81.4 & 81.3 & 80.9 & -0.1 \\
WR & 0.4 & 0.363 & 0.241 & 1.006 & 1.016 & 78.2 & 77.5 & 77.4 & 76.8 & -0.1 \\
WR & 0.6 & 0.351 & 0.256 & 1.006 & 1.022 & 74.2 & 73.4 & 73.4 & 72.5 & -0.1 \\
WR & 0.8 & 0.343 & 0.285 & 1.005 & 1.032 & 70.3 & 69.3 & 69.3 & 67.9 & -0.0 \\
\hdashline
NB & 0.0 & 0.151 & 0.229 & 1.013 & 1.062 & 85.9 & 86.1 & 85.9 & 83.8 & -0.2 \\
NB & 0.2 & 0.144 & 0.232 & 1.009 & 1.063 & 82.2 & 82.4 & 82.3 & 79.9 & -0.1 \\
NB & 0.4 & 0.136 & 0.241 & 1.006 & 1.066 & 78.6 & 78.5 & 78.5 & 75.9 & -0.1 \\
NB & 0.6 & 0.129 & 0.256 & 1.003 & 1.072 & 74.6 & 74.6 & 74.5 & 71.6 & -0.0 \\
NB & 0.8 & 0.122 & 0.285 & 0.999 & 1.082 & 70.7 & 70.4 & 70.4 & 67.0 & 0.0 \\
\bottomrule
\end{tabular}
\parbox{0.96\textwidth}{\footnotesize \textit{Note:} $\widehat{\Delta}$ denotes the plug-in effect size under $H_A$, and $\widehat{\tau}_{\Omega,A}$ denotes the plug-in overall tie probability under $H_A$. $P_A$ denotes the standard FORSS two-variance power calculation, $P_0$ denotes the FORSS null-variance calculation, and $P_{YG}$ denotes the Yu--Ganju/Barnhart tie-only calculation. The ratios $\widehat{\mathcal A}_0/\widehat{\mathcal A}_A$ and $\widehat{\mathcal A}_{YG}/\widehat{\mathcal A}_0$ quantify, respectively, the null-variance substitution relative to the full FORSS large-sample variance quantity and the tie-only approximation relative to the null large-sample variance quantity. Power values and $P_0-P_A$ are reported as percentages and percentage points, respectively. Emp. is the trial-level simulation benchmark.}
\end{table}

\begin{figure}[ht!]
\centering
\includegraphics[width=0.98\textwidth]{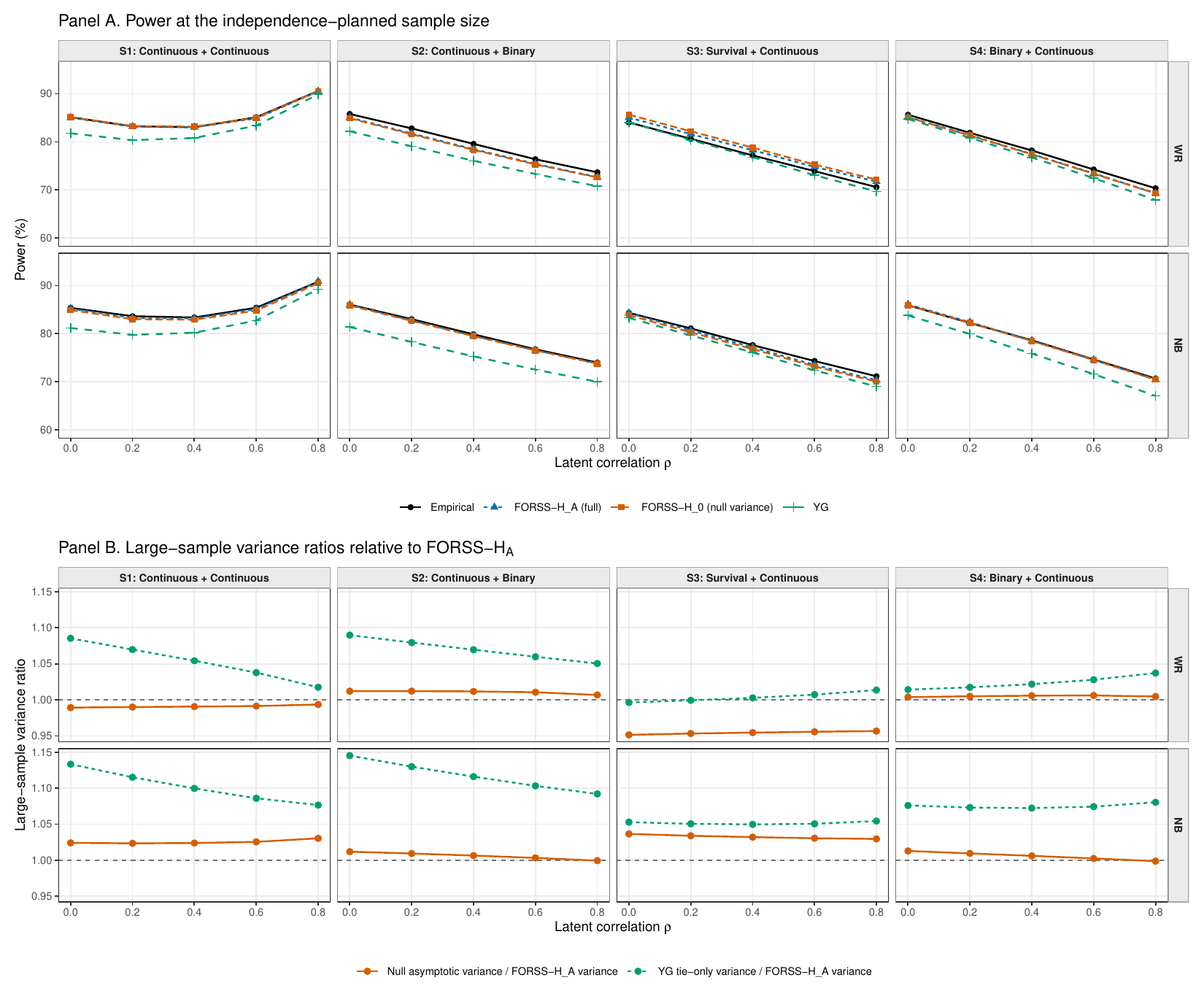}
\caption{Sensitivity analysis for the null-variance and tie-only variance approximations. Panel A compares empirical power with three calculated powers: the full FORSS two-variance calculation (FORSS-$H_A$), the null-variance approximation (FORSS-$H_0$), and the Yu--Ganju/Barnhart tie-only approximation at the sample size selected under $\rho=0$. Panel B displays the ratios $\widehat{\mathcal A}_0(\rho)/\widehat{\mathcal A}_A(\rho)$ and $\widehat{\mathcal A}_{YG}(\rho)/\widehat{\mathcal A}_A(\rho)$, which quantify the extent of the null-variance substitution and the tie-only approximation relative to the full FORSS large-sample variance quantity.}
\label{fig:appendix_equal_variance_sensitivity_all}
\end{figure}

The numerical results in Table~\ref{tab:appendix_equal_variance_sensitivity_all},
together with the graphical summary in
Figure~\ref{fig:appendix_equal_variance_sensitivity_all}, show that replacing the alternative variance by the null variance has a limited effect on the calculated power compared with the empirical power changes induced by increasing dependence. For WR, the FORSS-$H_0$ minus FORSS-$H_A$ power difference ranged from about $-0.1$ to 0.6 percentage points across all scenarios and correlations; for NB, the corresponding range was about $-0.4$ to 0.0 percentage points. Moreover, the direction was not uniformly conservative: the null-variance substitution increased WR power in S1 and S3, but decreased WR power in S2 and S4, while it generally decreased NB power. Thus, the large power losses seen under independence-based planning in S2--S4 cannot be attributed to the equal-variance approximation. They arise primarily because increasing dependence changes the overall win/loss contrast itself.

\subsection{Representative Comparison of Super-sample of 8,000 and 2,000 Runs}\label{app:nsp8000_vs_2000}

To provide a matched comparison of computational burden, we contrasted aligned runs with $N_{\textup{sp}}=8000$ against the standardized runs with $N_{\textup{sp}}=2000$ using one representative dependence setting per scenario, namely $\rho=0$. Both sets of runs used the same adaptive Monte Carlo thresholds, $\varepsilon_{\tau}=5\times 10^{-4}$ and $\varepsilon_{\xi}=10^{-4}$, and all computations used 20 cores.

Across all four scenarios, reducing $N_{\textup{sp}}$ from 8000 to 2000 substantially decreased peak worker memory usage from approximately 8--9~GB to about 0.8--0.9~GB. Under this matched-threshold comparison, the $N_{\textup{sp}}=2000$ runs were also faster in all four representative settings. Among the settings that converged for both super-population sizes, the largest time savings were seen in Scenarios~S2 and S4, whereas the difference was more modest in Scenario~S1. The runtime comparison for Scenario~S3 should be interpreted more cautiously: although the $N_{\textup{sp}}=2000$ run finished sooner, it reached $B_{\max}=3000$, while the aligned $N_{\textup{sp}}=8000$ run converged at $B_{\text{final}}=850$. Thus, the smaller super-population size is clearly attractive from a memory perspective and is often faster in practice, but the survival-plus-continuous setting remains the most sensitive to the reduction in $N_{\textup{sp}}$.

\begin{table}[ht!]
\centering
\caption{Representative computational comparison between aligned $N_{\textup{sp}}=8000$ runs and standardized $N_{\textup{sp}}=2000$ runs at $\rho=0$.}
\label{tab:nsp8000_vs_2000_rho0}
\setlength{\tabcolsep}{3.5pt}
\begin{tabular}{@{}l c ccc ccc@{}}
\toprule
& & \multicolumn{3}{c}{$N_{\textup{sp}}=8000$} & \multicolumn{3}{c}{$N_{\textup{sp}}=2000$} \\
\cmidrule(lr){3-5} \cmidrule(lr){6-8}
\textbf{Scenario} & $\rho$ & $b_{\text{final}}$ & Time (min) & Worker max (GB) & $B_{\text{final}}$ & Time (min) & Worker max (GB) \\
\midrule
S1: Cont + Cont & 0 & 250 & 12.3 & 7.80 & 900 & 6.5 & 0.80 \\
S2: Cont + Bin & 0 & 350 & 17.4 & 7.80 & 1350 & 7.5 & 0.80 \\
S3: Surv + Cont & 0 & 850 & 45.6 & 8.75 & 3000 & 14.3 & 0.91 \\
S4: Bin + Cont & 0 & 350 & 17.8 & 7.80 & 1400 & 6.4 & 0.81 \\
\bottomrule
\end{tabular}
\end{table}

\subsection{Additional Diagnostics for the HEART-FID Illustration}
\label{app:heartfid_computation}

This appendix provides additional computational and decomposition diagnostics for the HEART-FID design illustration. Table~\ref{tab:heartfid_computation} summarizes the adaptive Monte Carlo computation for the three dependence specifications considered in the HEART-FID illustration. All runs used the same super-population size $N_{\mathrm{sp}}=2000$ per arm, the same fixed design sample size $m=n=1,244$ per arm, and the same stopping parameters $b_{\min}=100$, $b_{\max}=3,000$, $\varepsilon_{\tau}=10^{-3}$, and $\varepsilon_{\xi}=10^{-4}$. All computations were carried out using 16 CPU cores in parallel. 

Overall, the adaptive estimator converged for all three dependence specifications before reaching the maximum number of batches. The required Monte Carlo budget varied across specifications, with the direct-input and calibrated dependence specifications requiring fewer batches than the independence specification in this example. These results suggest that the computational burden of the HEART-FID illustration is manageable for all three specifications under the chosen convergence criteria. Table~\ref{tab:heartfid_combined_scenarios} further decomposes the overall win probabilities into contributions from each level of the hierarchy. 

\begin{table}[ht!]
\centering
\caption{Adaptive Monte Carlo computational diagnostics for the HEART-FID illustration.}
\label{tab:heartfid_computation}
\setlength{\tabcolsep}{4pt}
\renewcommand{\arraystretch}{1.08}
\begin{tabular}{@{}l c c c c c c@{}}
\toprule
\textbf{Scenario} & $N_{\mathrm{sp}}$ & $b_{\text{final}}$ & Status & Time (min) & Max SE$(\tau)$ & Max SE$(\xi)$ \\
\midrule
$\bm{\mathcal R}^{\textup{ind}}$ & 2000 & 2000 & CONVERGED & 20.2 & $9.93\times 10^{-4}$ & $5.35\times 10^{-5}$ \\
$\bm{\mathcal R}^{\textup{dir}}$ & 2000 & 1300 & CONVERGED & 12.8 & $9.96\times 10^{-4}$ & $6.71\times 10^{-5}$ \\
$\bm{\mathcal R}^{\textup{cal}}$ & 2000 & 1550 & CONVERGED & 15.4 & $9.84\times 10^{-4}$ & $6.01\times 10^{-5}$ \\
\bottomrule
\end{tabular}
\parbox{0.92\textwidth}{\footnotesize \textit{Note:} All adaptive estimation runs used 16 CPU cores, $b_{\min}=100$, $b_{\max}=3,000$, $\varepsilon_{\tau}=10^{-3}$, and $\varepsilon_{\xi}=10^{-4}$. Runtime is reported in minutes from the adaptive estimation stage only. The fixed design used $m=n=1,244$ per arm, determined under the independent scenario using the WR.}
\end{table}

\begin{landscape}
\begin{table}
\centering
\caption{Win probability decomposition and overall win measures across three correlation scenarios in the HEART-FID trial.}
\label{tab:heartfid_combined_scenarios}
\begin{threeparttable}
\setlength{\tabcolsep}{3.0pt}
\renewcommand{\arraystretch}{1.08}
\footnotesize
\begin{tabular}{@{}l c c c ccc ccc ccc ccc cccc@{}}
\toprule
& \multicolumn{2}{c}{Correlation} & & \multicolumn{3}{c}{\textbf{Marginal $D_1$ (\%)}} & \multicolumn{3}{c}{\textbf{Cond.\ $D_2|D_1$ Tie (\%)}} & \multicolumn{3}{c}{\textbf{Cond.\ $D_3|D_1,D_2$ Tie (\%)}} & \multicolumn{3}{c}{\textbf{Overall Probs (\%)}} & \multicolumn{4}{c}{\textbf{Win Measures}} \\
\cmidrule(lr){2-3} \cmidrule(lr){5-7} \cmidrule(lr){8-10} \cmidrule(lr){11-13} \cmidrule(lr){14-16} \cmidrule(l){17-20}
Specification & Latent corr.\tnote{a} & Obs. corr.\tnote{b} & $\epsilon_3$ & Win & Loss & Tie & Win & Loss & Tie & Win & Loss & Tie & Win & Loss & Tie & WR & NB & WO & DOOR \\
\midrule
$\bm{\mathcal R}^{\textup{ind}}$ & (0.00, 0.00, 0.00) & (0.00, 0.00, 0.00) & 0 & 9.85 & 8.16 & 81.98 & 22.74 & 16.94 & 60.32 & 50.49 & 49.51 & 0.00 & 53.46 & 46.54 & 0.00 & 1.149 & 0.069 & 1.149 & 0.535 \\
$\bm{\mathcal R}^{\textup{dir}}$ & (-0.22, 0.52, -0.10) & (-0.15, 0.55, -0.06) & 0 & 9.87 & 8.18 & 81.96 & 21.65 & 16.31 & 62.05 & 50.11 & 49.89 & 0.00 & 53.09 & 46.91 & 0.00 & 1.132 & 0.062 & 1.132 & 0.531 \\
$\bm{\mathcal R}^{\textup{cal}}$ & (-0.30, 0.49, -0.17) & (-0.21, 0.53, -0.10) & 0 & 9.82 & 8.13 & 82.05 & 21.26 & 15.95 & 62.79 & 50.04 & 49.96 & 0.00 & 53.04 & 46.96 & 0.00 & 1.130 & 0.061 & 1.130 & 0.530 \\
\bottomrule
\end{tabular}
\begin{tablenotes}\footnotesize
\item[a] Latent Gaussian copula correlations: $(\rho_{12}, \rho_{13}, \rho_{23})$ for endpoint pairs $D_1$--$D_2$, $D_1$--$D_3$, and $D_2$--$D_3$.
\item[b] Observed pairwise associations in the simulated treatment-arm super-populations used by the FORSS adaptive estimator: $(\hat{a}_{12}, \hat{a}_{13}, \hat{a}_{23})$. Here $\hat{a}_{12}=2C\{D_1,D_2\}-1$ and $\hat{a}_{13}=2C\{D_1,D_3\}-1$, where $C$ is Harrell's concordance using administratively censored $D_1$; $\hat{a}_{23}$ is Kendall's tau-b between $D_2$ and $D_3$.
\item Note: WR = Win Ratio; NB = Net Benefit; WO = Win Odds; DOOR = Desirability of Outcome Ranking. $D_1$ = CV death (time-to-event); $D_2$ = HF hospitalizations (count); $D_3$ = 6MWD change (continuous). Cond. = conditional probability among pairs tied on all higher-priority endpoints.
\end{tablenotes}
\end{threeparttable}
\end{table}
\end{landscape}

\end{document}